\documentclass[twocolumn,english,reprint, superscriptaddress, breaklinks=true, showkeys, showpacs=false, nofootinbib]{revtex4}

\usepackage[T1]{fontenc}
\usepackage{color}
\usepackage{babel}
\usepackage{float}
\usepackage{amssymb}
\usepackage{listings} 
\usepackage{graphicx}
\usepackage{amsmath}
\usepackage{svg}
\usepackage{tikz}
\usepackage{braket}
\usepackage{cancel}

\usepackage[unicode=true,pdfusetitle, bookmarks=true,bookmarksnumbered=false,bookmarksopen=false, breaklinks=true,pdfborder={0 0 0},backref=false,colorlinks=true]{hyperref}
\usepackage{breakurl}
\definecolor{Code}{rgb}{0,0,0}
\definecolor{Decorators}{rgb}{0.5,0.5,0.5}
\definecolor{Numbers}{rgb}{0.5,0,0}
\definecolor{MatchingBrackets}{rgb}{0.25,0.5,0.5}
\definecolor{Keywords}{rgb}{0,0,1}
\definecolor{self}{rgb}{0,0,0}
\definecolor{Strings}{rgb}{0,0.63,0}
\definecolor{Comments}{rgb}{0,0.63,1}
\definecolor{Backquotes}{rgb}{0,0,0}
\definecolor{Classname}{rgb}{0,0,0}
\definecolor{FunctionName}{rgb}{0,0,0}
\definecolor{Operators}{rgb}{0,0,0}
\definecolor{Background}{rgb}{0.98,0.98,0.98}
\lstdefinelanguage{Python}{
numbers=left,
numberstyle=\footnotesize,
numbersep=1em,
xleftmargin=1em,
framextopmargin=2em,
framexbottommargin=2em,
showspaces=false,
showtabs=false,
showstringspaces=false,
frame=l,
tabsize=4,
basicstyle=\ttfamily\small\setstretch{1},
backgroundcolor=\color{Background},
commentstyle=\color{Comments}\slshape,
stringstyle=\color{Strings},
morecomment=[s][\color{Strings}]{"""}{"""},
morecomment=[s][\color{Strings}]{'''}{'''},
morekeywords={import,from,class,def,for,while,if,is,in,elif,else,not,and,or,print,break,continue,return,True,False,None,access,as,,del,except,exec,finally,global,import,lambda,pass,print,raise,try,assert},
keywordstyle={\color{Keywords}\bfseries},
morekeywords={[2]@invariant,pylab,numpy,np,scipy},
keywordstyle={[2]\color{Decorators}\slshape},
emph={self},
emphstyle={\color{self}\slshape},
}

\makeatletter
\@ifundefined{textcolor}{}
{%
 \definecolor{BLACK}{gray}{0}
 \definecolor{WHITE}{gray}{1}
 \definecolor{RED}{rgb}{1,0,0}
 \definecolor{GREEN}{rgb}{0,1,0}
 \definecolor{BLUE}{rgb}{0,0,1}
 \definecolor{CYAN}{cmyk}{1,0,0,0}
 \definecolor{MAGENTA}{cmyk}{0,1,0,0}
 \definecolor{YELLOW}{cmyk}{0,0,1,0}
}

\hypersetup{colorlinks=true,citecolor=blue,linkcolor=cyan,urlcolor=blue,filecolor= green, breaklinks=true}
\usepackage{url}
\usepackage{breakurl}

\newtheorem{theorem}{Theorem}
\newtheorem{corollary}{Corollary}
\newtheorem{lemma}{Lemma}

\makeatother

\begin{document}
\title{A Classical-Quantum Hybrid Architecture for Physics-Informed Neural Networks}
\author{Said Lantigua\href{https://orcid.org/0009-0008-4594-9312}{\includegraphics[scale=0.04]{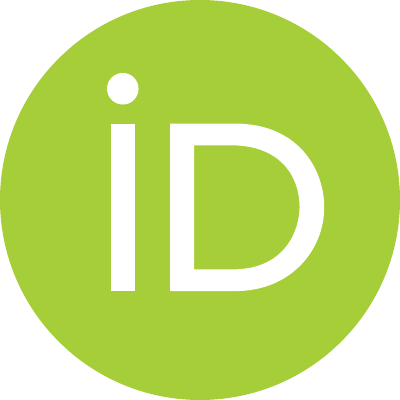}}}
\email{saidjose@lncc.br}
\affiliation{National Laboratory for Scientific Computing (LNCC), Av. Getúlio Vargas, 333 - Quitandinha, Petrópolis, RJ, 25651-075, Brazil}
\author{Gilson Giraldi\href{https://orcid.org/0000-0003-0623-9461}{\includegraphics[scale=0.04]{orcidid.pdf}}}
\email{gilson@lncc.br}
\affiliation{National Laboratory for Scientific Computing (LNCC), Av. Getúlio Vargas, 333 - Quitandinha, Petrópolis, RJ, 25651-075, Brazil}
\author{Renato Portugal\href{https://orcid.org/0000-0003-0894-4279}{\includegraphics[scale=0.04]{orcidid.pdf}}}
\email{portugal@lncc.br}
\affiliation{National Laboratory for Scientific Computing (LNCC), Av. Getúlio Vargas, 333 - Quitandinha, Petrópolis, RJ, 25651-075, Brazil}
\begin{abstract}
In this work, we introduce the Quantum-Classical Hybrid Physics-Informed Neural Network with Multiplicative and Additive Couplings (QPINN-MAC): a novel hybrid architecture that integrates the framework of Physics-Informed Neural Networks (PINNs) with that of Quantum Neural Networks (QNNs). Specifically, we prove that through strategic couplings between classical and quantum components, the QPINN-MAC retains the universal approximation property, ensuring its theoretical capacity to represent complex solutions of ordinary differential equations (ODEs). Simultaneously, we demonstrate that the hybrid QPINN-MAC architecture actively mitigates the barren plateau problem, regions in parameter space where cost-function gradients decay exponentially with circuit depth, a fundamental obstacle in QNNs that hinders optimization during training. Furthermore, we prove that these couplings prevent gradient collapse, ensuring trainability even in high-dimensional regimes. Thus, our results establish a new pathway for constructing quantum-classical hybrid models with theoretical convergence guarantees, which are essential for the practical application of QPINNs.
\end{abstract}
\maketitle
\section{Introduction}
\label{sec:1}
In recent years, the rapid advancement of classical machine learning has driven the development of sophisticated methods for modeling physical and chemical systems. Among these methods, Physics-Informed Neural Networks (PINNs) \cite{Raissi2019Physics, Karniadakis2021} stand out, integrating physical laws directly into the machine learning process. These networks enable the numerical solution of ordinary and partial differential equations (ODEs/PDEs) \cite{Demidovich,Piskunov197700,Piskunov197701,Makarenko198400} with theoretical convergence guarantees.

Furthermore, the effectiveness of PINNs has been demonstrated in various areas, such as fluid mechanics \cite{Cai2021, Sharma202300, Jin2021NSFnets}, elasticity \cite{Kamalia2023Elasticity}, smoke modeling \cite{ChuPhysicsInformed}, climatology \cite{Kashinath2021PhysicsInformed}, among others \cite{Karniadakis2021, Giraldi202300, Wu2018PhysicsInformed}. These networks, based on multilayer perceptron (MLP) architectures, are trained to minimize both the data fitting error and the residual of the governing equations \cite{pinns_overview}. This makes them a powerful framework for inverse and simulation problems where experimental data are scarce or noisy.

In parallel, the emergence of quantum computing has opened new frontiers for information processing \cite{nielsen_chuang}, with the potential to revolutionize the solution of computationally intractable problems. In this context, Quantum Physics-Informed Neural Networks (QPINNs) \cite{qpinn, qpinn_hybrid, Panichi2025QuantumPhysics, Markidis2022QuantumComputers, Trahan2024QuantumPINNs, ReddyVadyala202300} emerge as a natural extension of PINNs. They incorporate fundamental concepts of quantum mechanics, such as superposition, entanglement, and interference, to enhance the representational capacity of the PINN models \cite{vqa_survey,Peruzzo201400}.

This integration of quantum computing and machine learning also shows promise in areas such as computational fluid dynamics \cite{Sedykh202400}, quantum control \cite{Batra2024PhysicsInformed}, and solving linear differential equations \cite{Arrazola201900, Xiao202400, Xiao202401, Broeckmann202400, Siegl202400}.

However, this approach faces critical challenges, in particular the barren plateaus problem \cite{mcclean2018barren, cerezo2021cost, wang2023noise}. That is, regions in the parameter space where the gradients of the cost function decay exponentially with the number of qubits, becoming a problem that hinders the efficient training of deep quantum neural networks.

In this work, we introduce the Quantum-Classical Hybrid Physics-Informed Neural Network with Multiplicative and Additive Couplings (QPINN-MAC). This is an innovative hybrid architecture that combines the representational power of classical PINNs with quantum efficiency in a unified framework. Our approach employs multiplicative and additive couplings between the classical and quantum components, ensuring not only greater expressivity \cite{Paine202300, zhao2022hybrid, perez2020data} but also active mitigation of the barren plateaus problem.

Furthermore, we prove that QPINN-MAC satisfies a universal approximation theorem \cite{cybenko198900,hornik1991approximation, schuld2019quantum}, which guarantees its ability to represent solutions of ODEs with arbitrary accuracy. Moreover, we establish rigorous upper bounds for the gradient magnitude, proving that the architecture maintains computational efficiency even in high-dimensional regimes.

Thus, this manuscript is organized as follows: In Section \ref{sec:2}, we review related works and contextualize our contributions within the existing literature. In Section \ref{sec:3}, we present the detailed mathematical formulation of QPINN-MAC, including its hybrid structure and theoretical foundations, with demonstrations of the universal approximation theorem, gradient upper bound analysis, and barren plateaus mitigation. In Section \ref{sec:4}, we establish the universal approximation properties of our architecture. Finally, in Section \ref{sec:5}, we present conclusions and future perspectives.

\section{Related Works}
\label{sec:2}
Having introduced the QPINN-MAC hybrid architecture and its formal foundations, we now contextualize our contribution within the existing literature. This work operates at the intersection of two well-established frameworks: classical Physics-Informed Neural Networks (PINNs) and their quantum counterparts.

The first framework encompasses Physics-Informed Neural Networks (PINNs) \cite{Raissi2019Physics}. As referenced in Sections \ref{sec:1} and \ref{sec:3}, PINNs have introduced significant advances in numerical methods, proving effective for solving both direct and inverse problems across various disciplines \cite{Cai202100, Sharma202300, Giraldi202300, Jin2021NSFnets, Kamalia2023Elasticity, han2021solving}.

The second framework concerns Quantum Physics-Informed Neural Networks (QPINNs) \cite{goto2021universal, schuld2019quantum, zhao2022hybrid}, which emerge naturally from combining Quantum Neural Networks (QNNs) \cite{Beer2022QuantumNN, Schuld2021SupervisedQuantum} with quantum algorithms for solving differential equations, such as the Harrow-Hassidim-Lloyd algorithm \cite{Aram200900} and related approaches \cite{Berry201700, Clader201300}.

Theoretical foundations for quantum universal approximation have been established through recent breakthroughs. The universal approximation property for quantum models was first demonstrated by Goto et al. \cite{goto2021universal}, who proved that quantum machine learning models in quantum-enhanced feature spaces can approximate continuous functions arbitrarily well. Building on this, Gonon and Jacquier \cite{gonon2025universal} provided the first quantitative error bounds, showing that quantum neural networks with $\mathcal{O}(\varepsilon^{-2})$ weights and $\mathcal{O}(\lceil \log_2(\varepsilon^{-1}) \rceil)$ qubits can achieve approximation error $\varepsilon > 0$ for functions with integrable Fourier transforms. More recently, Manzano et al. \cite{Manzano2025Approximation} extended these results to Sobolev spaces $H^k$, demonstrating that parametrized quantum circuits can approximate functions in $L^p$, $C^0$, and $H^k$ under specific distances, with direct implications for differential equations applications. These theoretical advances provide the mathematical underpinning for the expressive power of quantum models in scientific computing.

In the realm of pioneering works on Quantum PINNs, early explorations by Markidis \cite{Markidis2022QuantumComputers}, Trahan et al. \cite{Trahan2024QuantumPINNs}, and Reddy-Vadyala and Betgeri \cite{ReddyVadyala202300} investigated the use of parametrized quantum circuits (PQCs) to represent or enhance solutions to differential equations. These hybrid models leverage the expressivity potential of QNNs to model complex functions, exploiting the advantages of quantum feature spaces \cite{goto2021universal, schuld2019quantum, zhao2022hybrid}. Recent work by Farea et al. \cite{farea2025qcpinn} has explicitly applied universal approximation theorems in the context of hybrid quantum-classical PINNs.

The development of quantum algorithms for differential equations has shown potential exponential accelerations for solving linear systems and, by extension, certain classes of differential equations \cite{Berry201400, Andrew202100, Linden202200, Montanaro201500, Childs201900, Cao201300, Arrazola201900}. Recent comparative studies by Broeckmann \cite{Broeckmann202400} and Xiao et al. \cite{Xiao202400, Xiao202401} have systematically analyzed quantum versus classical methods for differential equations. Despite their promise, these algorithms generally require fault-tolerant quantum computers and face substantial challenges in scalability and circuit complexity.

A fundamental challenge in this domain is the barren plateau problem \cite{mcclean2018barren, cerezo2021cost}, where the gradient of the loss function decays exponentially with the number of qubits and circuit depth, rendering optimization ineffective. Although previous studies have explored hybrid architectures to address this issue \cite{zhao2022hybrid, farea2025qcpinn}, most focus on empirical evidence or expressivity analyses without providing theoretical guarantees.

It is precisely within this context that our work distinguishes itself. Our proposal of the QPINN-MAC architecture emerges as a unique formulation that advances beyond prior approaches \cite{Sedykh202400, Paine202300, ReddyVadyala202300} through its mathematical rigor in demonstrating universality and, crucially, in proving that its gradient avoids exponential decay, being bounded above by $\mathcal{O}(1/\sqrt{\mathcal{N}N})$. This theoretical contribution solidifies the trainability guarantees of our model, representing a fundamental advance for the field of hybrid neural networks.
\section{Hybrid Model of Physics-Informed Neural Network with Multiplicative and Additive Couplings (QPINN-MAC)}
\label{sec:3}
First and foremost, it is essential to highlight that in this manuscript we use $d^{n}f/dt^{n}=f^{(n)}$ to denote the $n$-th order derivative with respect to time. We also define the sets $\mathcal{I}_{K} = \{1,2,3,\dots,K\}$, $\mathcal{I}_{M} = \{1,2,3,\dots,M\}$, $\mathcal{I}_{\mathcal{M}} = \{1,2,3,\dots,\mathcal{M}\}$, $\mathcal{I}_{N} = \{1,2,3,\dots,N\}$, $\mathcal{I}_{\mathcal{N}} = \{1,2,3,\dots\mathcal{N}\}$ and $\mathcal{I}_{S} = \{1,2,3,\dots,S\}$, where $K$, $M$, $\mathcal{M}$, $N$, $\mathcal{N}$, and $S$ are hyperparameters of the neural network architecture, representing respectively: the number of training points, output dimension, number of quantum circuit parameters, number of qubits, quantum circuit depth, and hidden layer dimension. The indicial notation, or Einstein notation, $\sum_{i\in\mathcal{I}_{N}}f_{i}f_{i}=f_{i}f_{i}$, where a repeated subscript indicates the sum over all components is applied along the text. Thus, we can write the spectral representation of an operator $\hat{\mathcal{A}}=\sum_{i,j \in\mathcal{I}_{N}}a_{ij}\ket{i}\bra{j}$, as $\hat{\mathcal{A}}=a_{ij}\ket{i}\bra{j}$ and consequently $\mbox{Tr}[\hat{A}]=\bra{s}[a_{ij}\ket{i}\bra{j}]\ket{s}=a_{ij}\delta_{si}\delta_{js} = a_{ii}$, for all $s \in \mathcal{I}_{S}$.

With this in mind, this section presents the construction of the hybrid model termed QPINN-MAC. That is, a formulation supported by the formalisms of PINNs \cite{Raissi2019Physics,Cheng2021DeepLearning,Kamalia2023Elasticity,Wu2018PhysicsInformed,Kashinath2021PhysicsInformed,Jin2021NSFnets,ChuPhysicsInformed,Vieira2021Anisotropic}, Parametrized Quantum Circuits (PQC) \cite{nielsen_chuang}, Quantum Neural Networks (QNN) \cite{Beer2022QuantumNN,Schuld2021SupervisedQuantum}, Hybrid Quantum Circuits (HQC) \cite{Trahan202400,ReddyVadyala202300,Beer202200,Qiskit202400} and QPINNs \cite{farea2025qcpinn,Markidis2022QuantumComputers,Panichi2025QuantumPhysics,Trahan2024QuantumPINNs,Batra2024PhysicsInformed}.

The formulation we present here will allow, in Section \ref{sec:3}, a reformulation of the Universal Approximation Theorem \cite{Haykin1999neural}. This will allow us to contextualize and extend the validity of this theorem to the QPINN-MAC architecture, which allows us to prove that such an architecture actively mitigates the barren plateaus problem.

Although PINNs can be applied to ordinary differential equations (ODEs) and partial differential equations (PDEs), this work focuses on the former. In this case, besides the ODEs, the initial conditions are included as new terms in the loss function.

For this reason, let us consider the ODE
\begin{equation}\label{eq:1}
    y^{\left(1\right)}=\mathcal{F}(t,y),
\end{equation}
where we define the vector function $y$ as:
\begin{eqnarray}\label{eq:2}
    y : \mathbb{R} & \longrightarrow & \mathbb{R}^{M} \\
    (t) & \longmapsto & y(t) = \left(y_{1}(t),\dots,y_{M}(t)\right), \nonumber
\end{eqnarray}
whose initial condition is given by:
\begin{equation}\label{eq:3}
    y(t_{0}) = y_{0}.
\end{equation}

By keeping the focus on the PINN formalism, two elements must be defined: a neural network and a physics-informed loss function. In the context of PINNs, the neural network is a multilayer perceptron (MLP), illustrated in Figure \ref{fig:MLP-Archtect}, composed of one input node (time $t$) and $M$ output nodes. The number of hidden layers $L-1$, the number of neurons per hidden layer, which is arbitrary and can vary between layers without depending on the configuration of the previous layers, as well as the activation function $\sigma$ are hyperparameters that must be specified by trial and error or using automated machine learning (AutoML) approaches \cite{feurer2015efficient,balaji2018benchmarking,snoek2012practical,kotthoff2017auto,elshawi2019automated}. In this architecture, $\mathcal{N_{P_{C}}}$ is the number of weights and biases.


\begin{widetext}
    \begin{center}
        \begin{figure}[!ht]
            \centering
            \includegraphics[width=0.90\linewidth]{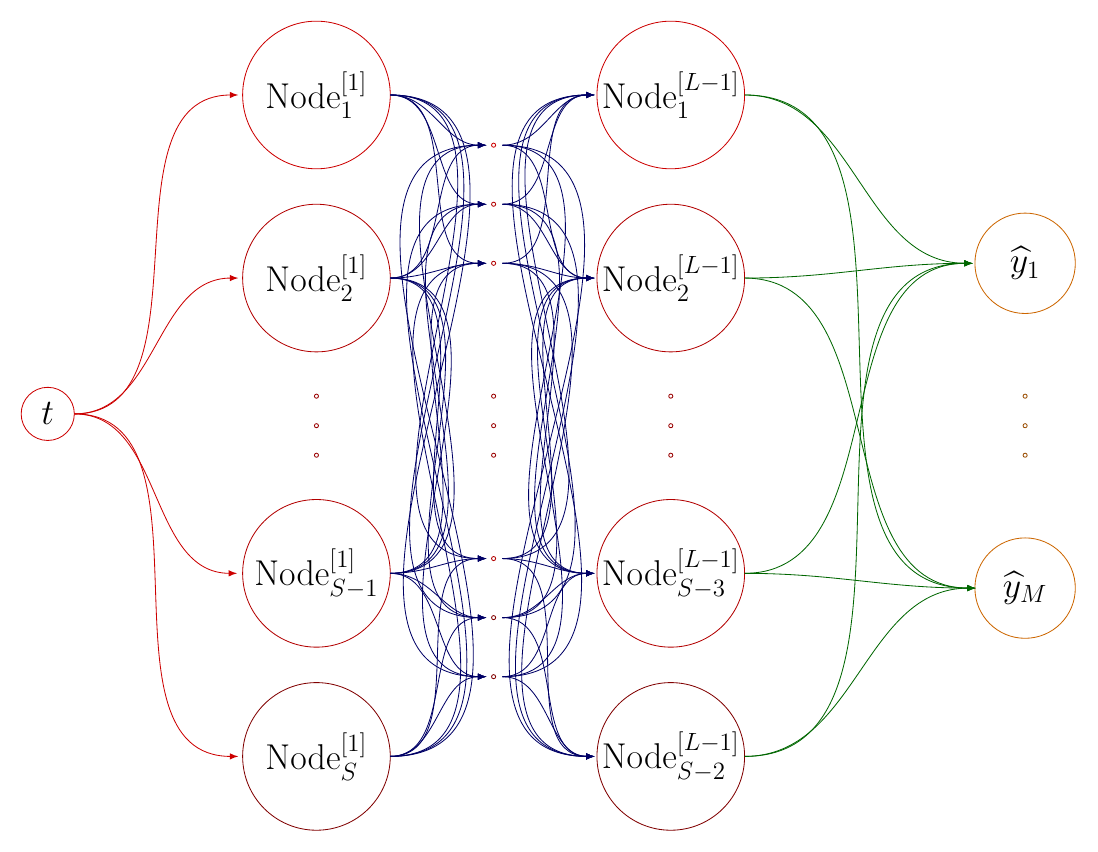}
            \caption{Architecture of the Multilayer Perceptron (MLP) used in the Physics-Informed Neural Networks (PINNs) framework. The classical neural network supports multiple hidden layers (minimum of one) with flexible layer sizes, where the number of neurons in each layer can be chosen independently without architectural constraints.}
            \label{fig:MLP-Archtect}
        \end{figure}
    \end{center}
\end{widetext}
Therefore, let $\vec{\mathcal{W}}$ be the set of weights and biases of the neural network and given a time $t$, the network computes a map
\begin{widetext}
    \begin{eqnarray}\label{eq:4}
        \widehat{y}\left(\cdot,\vec{\mathcal{W}}\right) : \mathbb{R} & \longrightarrow & \mathbb{R}^{M} \\
        (t) & \longmapsto & \widehat{y}\left(t,\vec{\mathcal{W}}\right) = \left(\widehat{y}_{1}\left(t,\vec{\mathcal{W}}\right),\widehat{y}_{2}\left(t,\vec{\mathcal{W}}\right),\ldots,\widehat{y}_{M}\left(t,\vec{\mathcal{W}}\right)\right). \nonumber
    \end{eqnarray}
\end{widetext}
Then, to find the optimal $\vec{\mathcal{W}^{\ast}}$ that provides a map $\widehat{y}\left(\cdot,\vec{\mathcal{W}^{\ast}}\right)$ that approximates the solution of the problem (\ref{eq:1})-(\ref{eq:3}), we need a loss function that connects the system dynamics and the network weight optimization. This is implemented using equations (\ref{eq:1}), together with the initial conditions (\ref{eq:3}) and, possibly, some known solutions $y(t_{k})$, $k\in\mathcal{I}_{K}$, to define the composite loss function as:
\begin{equation}\label{eq:5}
    \mathfrak{L}_{ODECS}=\mathfrak{L}_{IC}+\mathfrak{L}_{ODE}+\mathfrak{L}_{SOL}.
\end{equation}
Where
\begin{widetext}
    \begin{eqnarray}
        \mathfrak{L}_{IC} &=& \displaystyle {\left\Vert \widehat{y}(t_{0};\vec{\mathcal{W}})-y_{0}\right\Vert^{2}}, \label{eq:6} \\
        \mathfrak{L}_{ODE} &=& \displaystyle \sum_{k\in\mathcal{I}_{K}}\left\Vert \widehat{y}^{\left(1\right)}(t_{k},\vec{\mathcal{W}})-\mathcal{F}(t_{k},\widehat{y}(t_{k},\vec{\mathcal{W}})\right\Vert^{2}, \label{eq:7} \\
        \mathfrak{L}_{SOL} &=& \displaystyle {\sum_{\substack{k\in\mathcal{I}_{K}}}\left\Vert\widehat{y}(t_{k};\vec{\mathcal{W}})-y(t_{k})\right\Vert^{2}}. \label{eq:8}
    \end{eqnarray}
\end{widetext}
Furthermore, regarding the MLP, it is important to remember that each neuron (perceptron) receives input signals, performs a mathematical operation on them, and passes the result to the neurons in the next layer, according to the expression:
\begin{widetext}
    \begin{equation}\label{eq:9}
        x_{j}^{[l]}(t) = \sigma \biggl(\mathcal{W}_{ij}^{[l]}x_{i}^{[l-1]}(t)+b_{j}^{[l]}\biggr),\quad \mbox{with} \quad i \in \mathcal{I}_{S},j \in \mathcal{I}_{M},
    \end{equation}
\end{widetext}
where $x_{j}^{[l]}$ is the output of the $j$-th neuron in the $l$-th layer, $x_{i}^{[l-1]}$ is the output of the $i$-th neuron in the $[l-1]$-th layer, $\mathcal{W}_{ij}^{[l]}$ is the connection weight between the $i$-th neuron of the $[l-1]$-th layer and the $j$-th neuron of the $l$-th layer, $b_{j}^{[l]}$ is the bias term of the $j$-th neuron in the $l$-th layer, and $\sigma$ is an activation function \cite{Haykin1999neural}. Consequently, if the output layer of the MLP has index $l=L$, then the neural network output is given by:
\begin{widetext}
    \begin{equation}\label{eq:10}
        \widehat{y}_{j}\left(t,\vec{\mathcal{W}}\right)=\sigma\biggl(\mathcal{W}_{ij}^{[L]}x_{i}^{[L-1]}(t)+b_{j}^{[L]}\biggr),\quad \mbox{with} \quad i \in \mathcal{I}_{S},j \in \mathcal{I}_{M}.
    \end{equation}
\end{widetext}

Now, it is important to emphasize that the MLP's ability to learn complex tasks, such as classification and regression, is derived from adjusting the weights $\mathcal{W}_{ij}^{[l]}$ during training. However, in the case of PINNs, this process involves optimizing the loss expression (\ref{eq:5}) with respect to the neural network weights.

Continuing the theoretical foundation of this work, it is essential to define the concept of quantum nodes and variational layers, illustrated in Figure \ref{fig:QNode}. Hence, we observe that a quantum node can incorporate multiple variational layers responsible for applying rotations to the qubits with parameters $\theta_{k}^{j}$ to be optimized during training. Furthermore, performing measurements on the qubits of each quantum node is a necessary step to allow its connection with the other parts of the neural network, as also indicated in Figure \ref{fig:QNode}.

Generally, in each quantum variational layer there is a set of quantum gates  $\{\hat{U}_{k}^{j}(\theta_{k}^{j}), \; \theta_{k}^{j} \in \mathbb{R}\}_{\substack{k\in\mathcal{I}_{N} \\j\in\mathcal{I}_{\mathcal{N}}}}$, where $\theta_{k}^{j}$ compose a set of adjustable parameters:  Then, through a parameter vector
\begin{equation}\label{eq:12}
    \vec{\theta}^{j} = \left(\theta_{1}^{j},\dots,\theta_{N}^{j} \right),
\end{equation}
we define the operator $\mathcal{U}$ as:
\begin{widetext}
    \begin{equation}\label{eq:13}
        \begin{split}
            \mathcal{U}(\vec{\theta}^{j}) &= \mathcal{U} \left(\theta_{1}^{j},\dots,\theta_{N}^{j} \right) = \hat{U}_{N}^{j} \left( \theta_{N}^{j} \right) \otimes \dots \otimes \hat{U}_{1}^{j} \left( \theta_{1}^{j} \right),
        \end{split}
    \end{equation}
\end{widetext}
which acts on the Hilbert space $\mathcal{H}_{N}=\mbox{span}\{\ket{i_{0}}\otimes\dots\otimes\ket{i_{N-1}}\}_{i_{k}\in\{0,1\}}$. Because of this, it is possible to define a universal parametrized quantum circuit (PQC), whose quantum state in the space $\mathcal{H}_{N}$ is given by:
\begin{widetext}
    \begin{eqnarray}\label{eq:14}
        \ket{\widetilde{\Psi}(\vec{\theta}^{1})} &=& \mathcal{U}\left( \theta_{1}^{1},\dots,\theta_{N}^{1} \right) \ket{0 \dots 0} \\
        &=& \hat{U}_{N}^{1} \left( \theta_{N}^{1} \right) \otimes \dots \otimes \hat{U}_{1}^{1} \left( \theta_{1}^{1} \right) \ket{0} \otimes \dots \otimes \ket{0}. \nonumber
    \end{eqnarray}
\end{widetext}
In this sense, by choosing each $\hat{U}_{k}^{j}(\theta_{k}^{j})$ as a rotation gate $R_{Y}$, combining them with Hadamard gates $H$ and the conditional phase gate $CP(\phi)$ we obtain the $1$-th variational layer, \(\mathcal{VL}^{1}\), defined by:
\begin{widetext}
    \begin{equation}\label{eq:15}
        \ket{\Psi(\vec{\theta}^{1})} = \mathcal{VL}^{1} (\vec{\theta}^{1})\ket{0} \otimes \dots \otimes \ket{0} = CP(\phi) \left[ \left[ H R_{Y} \left( \theta_{1}^{1} \right) \right] \otimes R_{Y} \left( \theta_{2}^{1} \right) \otimes \dots \otimes R_{Y} \left( \theta_{N}^{1} \right) \right] \ket{0} \otimes \dots \otimes \ket{0}.
    \end{equation}
\end{widetext}
Where $\phi \in [0,2\pi]$ and the gates are defined as
\begin{widetext}
    \begin{eqnarray}
        H &=& \left[ (-1)^{\gamma \nu}/\sqrt{2} \right] \ket{\gamma} \bra{\nu}, \label{eq:16} \\
        R_{Y}(\theta_{k}^{j}) &=& \cos{\left(\theta_{k}^{j}/2 \right)} \openone + \sin{\left(\theta_{k}^{j}/2 \right)} \left[ \ket{1}\bra{0} - \ket{0} \bra{1} \right], \label{eq:17} \\
        CP(\phi) &=& \openone \otimes \dots \otimes \openone + [\exp{(-i \phi)} - 1] \ket{1} \bra{1} \otimes \dots \otimes \ket{1} \bra{1}, \label{eq:18} \\
        \openone &=& \ket{0} \bra{0} + \ket{1} \bra{1}, \label{eq:19}
    \end{eqnarray}
\end{widetext}
with $i=\sqrt{-1}$, $\gamma = 0,1$, $\nu = 0,1$, $k \in \mathcal{I}_{N}$ and $j \in \mathcal{I}_{\mathcal{N}}$. Consequently, by successively applying $\mathcal{N}$ variational layers to the $N$ qubits of the quantum circuit, we obtain the general state of the QNode given by:
\begin{widetext}
    \begin{equation}\label{eq:20}
        \ket{\Psi(\vec{\Theta})} = \ket{\Psi(\vec{\theta}^{1},\dots,\vec{\theta}^{\mathcal{N}})} = \left[ \displaystyle{\prod_{\substack{j \in\mathcal{I}_{\mathcal{N}}}} \mathcal{VL}^{j} (\vec{\theta}^{j}) } \right] \ket{0 \dots 0}.
    \end{equation}
\end{widetext}
Therefore, given the set $\mathbb{O}$ of all quantum observables defined in $\mathcal{H}_N$, the expected value of an observable $\hat{O}\in \mathbb{O}$ for a given $\ket{\Psi(\vec{\Theta})}$ is given by:
\begin{widetext}
    \begin{eqnarray}\label{eq:21}
        \braket{\hat{O}}_{\vec{\Theta}} &=& \bra{\Psi(\vec{\Theta})} \hat{O} \ket{\Psi(\vec{\Theta})} \\
        &=& \bra{0 \dots 0} \left[ \displaystyle{\prod_{\substack{j \in\mathcal{I}_{\mathcal{N}}}} \mathcal{VL}^{j} (\vec{\theta}^{j}) } \right]^{\dagger} \hat{O} \left[ \displaystyle{\prod_{\substack{j \in\mathcal{I}_{\mathcal{N}}}} \mathcal{VL}^{j} (\vec{\theta}^{j}) } \right] \ket{0 \dots 0}, \nonumber
    \end{eqnarray}
\end{widetext}
whose classically optimized value allows parameter adjustment \cite{Trahan202400,ReddyVadyala202300,Beer202200,Qiskit202400,nielsen_chuang}.
\begin{widetext}
    \begin{center}
        \begin{figure}[!ht]
            \centering
            \includegraphics[width=0.90\linewidth]{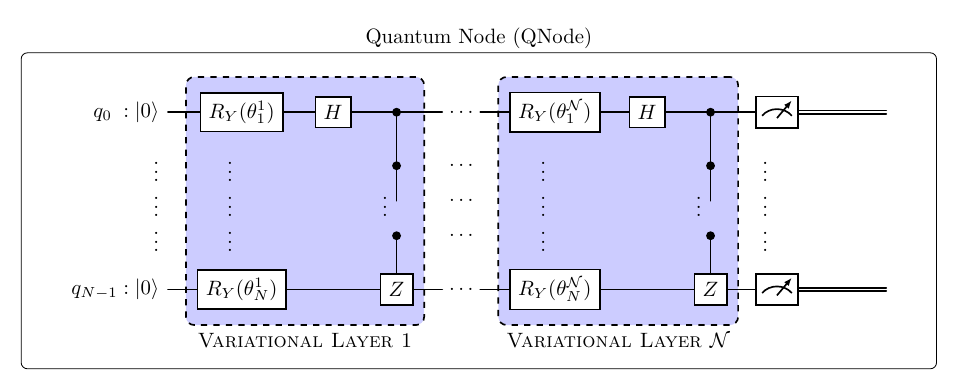}
            \caption{Architecture of a quantum neural network node (QNode) composed of $\mathcal{N}$ variational layers. Each layer consists of: (i) rotation gates $R_{Y}\left( \theta_{k}^{j} \right)$ applied to each qubit, (ii) a Hadamard gate ($H$) on the first qubit, (iii) a conditional phase gate $CP(\phi)$ with $\phi = \pi$ (equivalent to a controlled $Z$ gate) acting on the $N$-th qubit conditioned on the others, and (iv) measurements at the end of the circuit. The number of qubits and layers is configurable.}
            \label{fig:QNode}
        \end{figure}
    \end{center}
\end{widetext}
Thus, we already have the necessary elements to introduce the basic structure of our Quantum-Classical Hybrid architecture, which is illustrated in Figure \ref{fig:Simplified-Architect}. Specifically, this architecture integrates two main coupling schemes, the first of which is the multiplicative coupling scheme with output function given by the multiplicative form $y_{MC}(t) \propto Out(t)\braket{\hat{O}}_{\vec{\Theta}}$.

The second one is the additive coupling scheme with output function given by the additive form $y_{AC}(t) \propto Out(t)+\braket{\hat{O}}_{\vec{\Theta}}$. Therefore, the non-trivial combination of these schemes, which we will present shortly, generates the output function $y_{MAC}(t) \propto Out(t)\braket{\hat{O}}_{\vec{\Theta}} + \braket{\hat{O}}_{\vec{\Theta}}$.
\begin{widetext}
    \begin{center}
        \begin{figure}[!ht]
            \centering
            \includegraphics[width=0.90\linewidth]{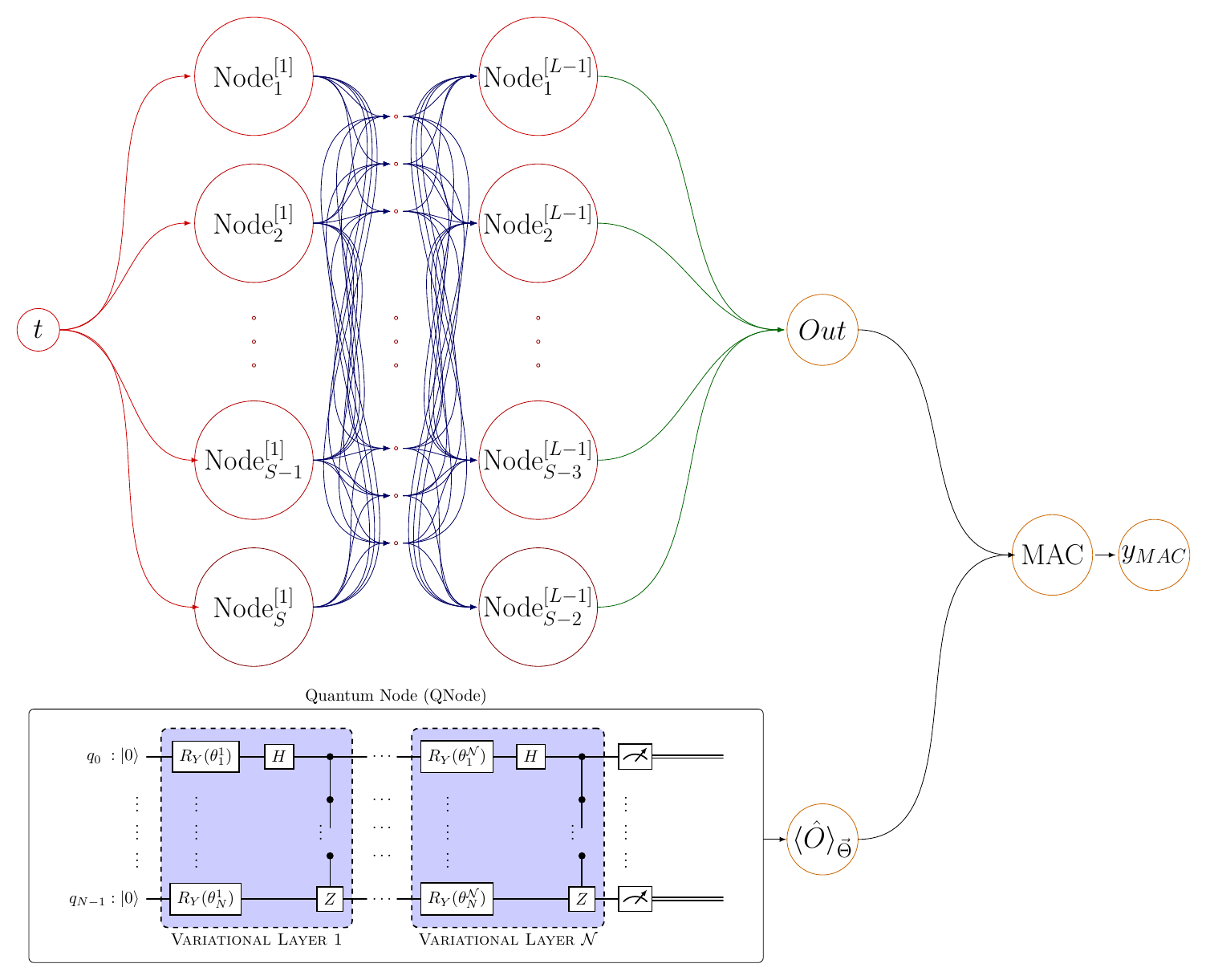}
            \caption{Scheme of the Quantum-Classical Hybrid Architecture with $MAC$ (multiplicative/additive) coupling modes between classical ($Out$) and quantum ($\braket{\hat{O}}_{\vec{\Theta}}$) outputs. The classical neural network admits multiple hidden layers (minimum of one) and an arbitrary number of neurons per layer, independent of the configuration of the other layers.}
            \label{fig:Simplified-Architect}
        \end{figure}
    \end{center}
\end{widetext}
However, the simplified coupling scheme previously discussed proves insufficient for generating outputs that accurately approximate the solution to the boundary value problem described by equations (\ref{eq:1})-(\ref{eq:3}). The fundamental limitation arises from the scalar nature of the preliminary architecture, which cannot capture the multidimensional character of the solution vector $y(t) \in \mathbb{R}^M$.

To address this critical constraint and enable the practical implementation of the QPINN-MAC model, we introduce an enhanced architectural paradigm where the classical neural network component generates $M$ distinct outputs, each corresponding to one component of the solution vector.

Crucially, each of these $M$ outputs interfaces with a dedicated quantum node (QNode) through the multiplicative and additive coupling ($MAC$) mechanism, establishing a one-to-one correspondence between solution dimensions and quantum processing units.

This refined configuration, comprehensively illustrated in Figure \ref{fig:QPINN-MAC-Architect}, establishes a sophisticated hybrid computing framework where classical and quantum processing collaborate synergistically across multiple dimensions. The architecture ensures that each component $y_j(t)$ of the solution vector benefits from both the robust pattern recognition capabilities of classical neural networks and the enhanced representational power of quantum circuits.

Through this distributed processing approach, the network achieves the necessary expressive capacity to adequately approximate the complete solution of the coupled differential equations (\ref{eq:1})-(\ref{eq:3}), while simultaneously maintaining the trainability advantages afforded by the MAC coupling scheme.

The modular design also permits independent optimization of each quantum-classical pathway, enabling tailored representations for different components of the solution dynamics.

\begin{widetext}
    \begin{center}
        \begin{figure}[!ht]
            \centering
            \includegraphics[width=0.90\linewidth]{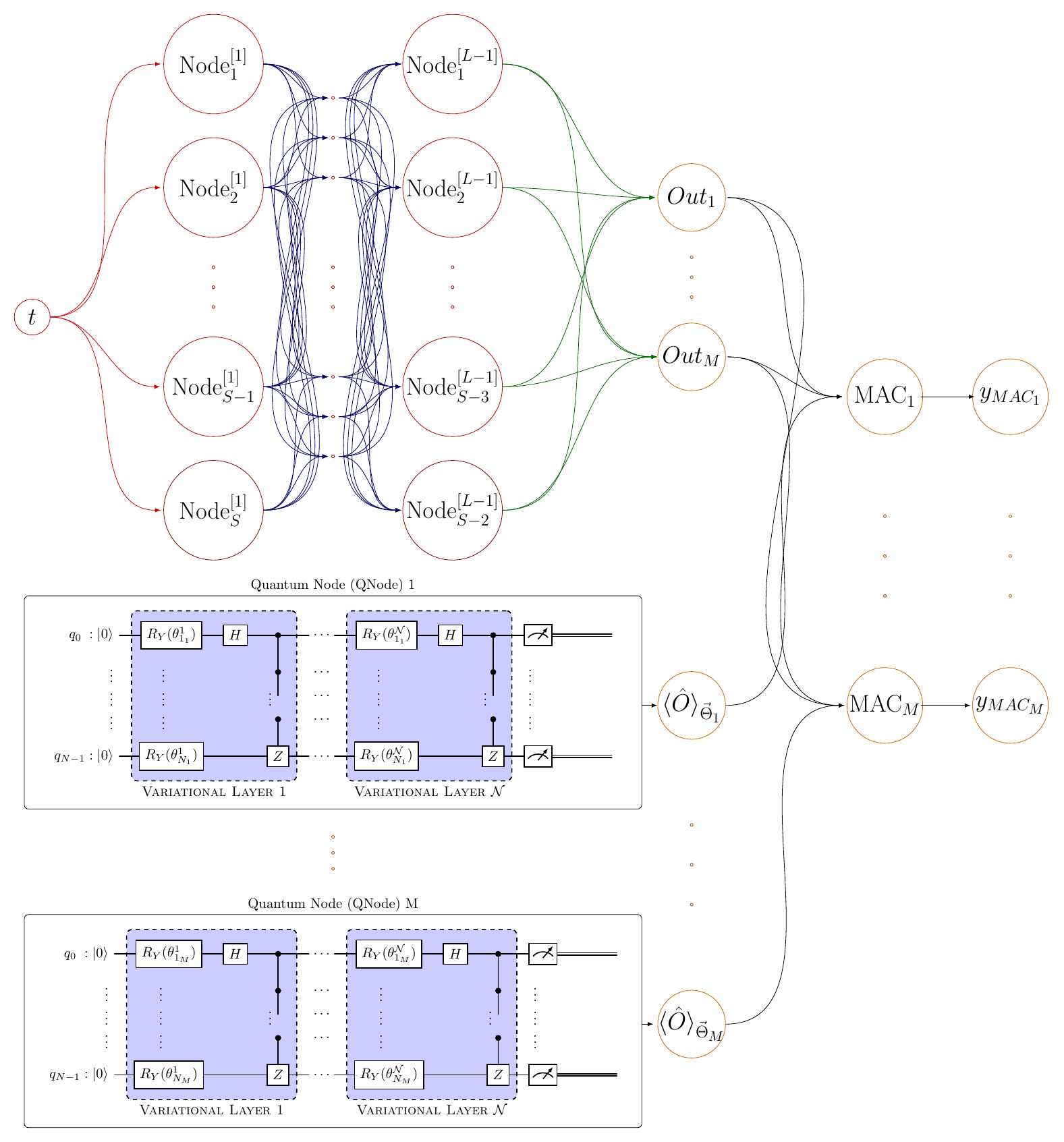}
            \caption{Proposed architecture of the QPINN-MAC model (Quantum Physics-Informed Neural Network with Multiplicative/Additive Coupling), illustrating the integration of $M$ quantum nodes (QNodes) coupled to the $M$ outputs of the classical neural network via $MAC$ modes. The classical network has a flexible architecture: multiple hidden layers (minimum of one) and a variable number of neurons per layer.}
            \label{fig:QPINN-MAC-Architect}
        \end{figure}
    \end{center}
\end{widetext}

Therefore, given the set of weights $\vec{\mathcal{W}}$ and biases of the classical neural network, the set of parameters
\begin{equation}\label{eq:}
    \vec{\Xi} = \left( \vec{\Theta}_{1},\dots,\vec{\Theta}_{M} \right)
\end{equation}
and a time $t$, the network computes a map
\begin{widetext}
    \begin{eqnarray}\label{eq:22}
        y_{MAC}\left(\hat{O},\cdot,\vec{\mathcal{W}},\vec{\Xi}\right) : \mathbb{R} & \longrightarrow & \mathbb{R}^{M} \\
        (t) & \longmapsto & y_{MAC}\left(\hat{O},t,\vec{\mathcal{W}},\vec{\Xi}\right) = \left(y_{MAC_{1}}\left(\hat{O},t,\vec{\mathcal{W}},\vec{\Theta}_{1}\right),\dots,y_{MAC_{M}}\left(\hat{O},t,\vec{\mathcal{W}},\vec{\Theta}_{M}\right)\right). \nonumber
    \end{eqnarray}
\end{widetext}

Thus, the functions $y_{MAC_{j}}(\hat{O},t,\vec{\mathcal{W}},\vec{\Theta}_{j})$ of the QPINN-MAC outlined in Figure \ref{fig:Simplified-Architect} is obtained through the encoding:
\begin{widetext}
    \begin{equation}\label{eq:23}
        y_{MAC_{j}}(\hat{O},t,\vec{\mathcal{W}},\vec{\Theta}_{j}) = \mbox{Tr}\left[\rho(t,\vec{\mathcal{W}},\vec{\Theta}_{j})\left(\hat{A}_{\mathcal{C}}(t,\vec{\mathcal{W}})\otimes\hat{O}\right)\right].
    \end{equation}
\end{widetext}
In expression (\ref{eq:23}), the term $\rho(t,\vec{\mathcal{W}},\vec{\Theta}_{j})$ denotes the composite density matrix of the system. It is obtained by the tensor product of the classical density matrix, $\rho_{\mathcal{C}}(t,\vec{\mathcal{W}})$, which contains the classical information of the system, and the quantum density matrix, $\rho_{\mathcal{Q}}(\vec{\Theta}_{j})$, corresponding to the general state of the QNode $\ket{\Psi(\vec{\Theta}_{j})}$. Thus, we have $\rho(t,\vec{\mathcal{W}},\vec{\Theta}_{j})=\rho_{\mathcal{C}}(t,\vec{\mathcal{W}})\otimes\rho_{\mathcal{Q}}(\vec{\Theta}_{j})$, with
\begin{equation}\label{eq:24}
    \rho_{\mathcal{C}}(t,\vec{\mathcal{W}}) = \mathcal{C}_{lm}(t,\vec{\mathcal{W}})\ket{l}\bra{m},
\end{equation}
\begin{equation}\label{eq:25}
    \rho_{\mathcal{Q}}(\vec{\Theta}_{j}) = \ket{\Psi(\vec{\Theta}_{j})}\bra{\Psi(\vec{\Theta}_{j})},
\end{equation}
where the functions
\begin{eqnarray}\label{eq:26}
    \mathcal{C}_{lm}(\cdot,\vec{\mathcal{W}}): R & \longrightarrow & R, \;\; l,m \in \mathcal{I}_{N}
\end{eqnarray}
encode the classical information of the network outlined in figure \ref{fig:QPINN-MAC-Architect} and the states $\{\ket{l}\}_{l \in \mathcal{I}_{N}}$ form an orthonormal basis in a Hilbert space $\mathcal{H}_{\mathcal{C}} = \mbox{span}\{\ket{l}\}_{l \in \mathcal{I}_{N}}$. Furthermore, in this Hilbert space $\mathcal{H}_{C}$ it is possible to define the following operators:
\begin{equation}\label{eq:27}
    \hat{A}_{\mathcal{C}}(t,\vec{\mathcal{W}}) = \hat{A}_{\mathcal{C}}^{\prime}(t,\vec{\mathcal{W}})+\openone,
\end{equation}
with $\hat{A}_{C}^{\prime}$ given by:
\begin{equation}\label{eq:28}
    \hat{A}_{\mathcal{C}}^{\prime}(t,\vec{\mathcal{W}}) = \mathcal{C}_{l^{\prime}m^{\prime}}^{\prime}(t,\vec{\mathcal{W}})\ket{l^{\prime}}\bra{m^{\prime}},
\end{equation}
where
\begin{eqnarray}\label{eq:29}
    \mathcal{C}_{l^{\prime}m^{\prime}}^{\prime}(\cdot,\vec{\mathcal{W}}): R & \longrightarrow & R, \;\; l^{\prime}, m^{\prime} \in \mathcal{I}_{N}
\end{eqnarray}
are functions that also encode the classical information of the network outlined in figure \ref{fig:QPINN-MAC-Architect}.

In turn, the observable $\hat{O}$ can be expressed as a sum of Pauli-Z operators acting on different qubits, according to:
\begin{widetext}
    \begin{equation}\label{eq:30}
        \hat{O} = \underbrace{\hat{Z}\otimes \openone \otimes \dots \otimes \openone}_{\text{N-operators}}+\underbrace{\openone\otimes\hat{Z}\otimes\openone\otimes\dots\otimes\openone}_{\text{N-operators}}+\dots+\underbrace{\openone\otimes\dots\otimes\openone\otimes\hat{Z}}_{\text{N-operators}}.
    \end{equation}
\end{widetext}

Therefore, substituting expressions (\ref{eq:24}), (\ref{eq:25}), (\ref{eq:27}) and (\ref{eq:28}) into (\ref{eq:23}), and remembering the property that the trace of a density matrix is $\mbox{Tr}[\rho_{\mathcal{C}}(t,\vec{\mathcal{W}})]=1$, it is possible to show that
\begin{widetext}
    \begin{eqnarray}\label{eq:31}
        y_{MAC_{j}}(\hat{O},t,\vec{\mathcal{W}},\vec{\Theta}_{j}) &=& \left[\mathcal{C}_{nm}(t,\vec{\mathcal{W}})\mathcal{C}_{mn}^{\prime}(t,\vec{\mathcal{W}}) + 1\right]\braket{\hat{O}}_{\vec{\Theta}_{j}} \\
        &=& \left[Out_{j}(t,\vec{\mathcal{W}}) + 1\right]\braket{\hat{O}}_{\vec{\Theta}_{j}}, \nonumber
    \end{eqnarray}
\end{widetext}
with $\mathcal{C}_{nm}(t,\vec{\mathcal{W}})\mathcal{C}_{mn}^{\prime}(t,\vec{\mathcal{W}}) = Out_{j}(t,\vec{\mathcal{W}})$. Then, it suffices to make the appropriate encoding in (\ref{eq:31}) so that $Out_{j}(t,\vec{\mathcal{W}}) = \widehat{y}_{j}(t,\vec{\mathcal{W}})$, and thus obtain the expression
\begin{widetext}
    \begin{equation}\label{eq:32}
        y_{MAC_{j}}(\hat{O},t,\vec{\mathcal{W}},\vec{\Theta}_{j}) = \widehat{y}_{j}(t,\vec{\mathcal{W}})\braket{\hat{O}}_{\vec{\Theta}_{j}} + \braket{\hat{O}}_{\vec{\Theta}_{j}}, \; j \in \mathcal{I}_{M}.
    \end{equation}
\end{widetext}

It is important to note that, if in this approach we consider $\braket{\hat{O}}_{\vec{\Theta}_{j}} \approx \mbox{constant}$, then a classical regime dominates. On the other hand, if $\widehat{y}_{j}(t,\vec{\mathcal{W}}) \approx 0$ for all $j\in\mathcal{I}_{M}$, a quantum regime predominates.

Therefore, any of these configurations would lead to a less expressive approach, reducing to those already considered in the literature \cite{Markidis2022QuantumComputers, Trahan2024QuantumPINNs, Raissi2019Physics, farea2025qcpinn}. The flexibility offered by the combination of classical and quantum regimes is crucial for the model's expressivity.

Thus, the formulation of expression (\ref{eq:22}) with the components defined by (\ref{eq:32}) culminates in the proposed QPINN-MAC model, whose architecture is detailed in Figure \ref{fig:QPINN-MAC-Architect}. Thereby, we complete the description of the architecture and operational principles of QPINN-MAC, highlighting the coupling forms and operation regimes.

Once the hybrid architecture proposed in this work is established, we proceed, in the following section, to the theoretical foundation of the model's universal approximation properties.

\section{Universal Approximation Properties of QPINN-MAC}
\label{sec:4}
Having introduced the hybrid QPINN-MAC architecture, we now proceed to construct the formal proof of its universal approximation capabilities. To ground our analysis, it is crucial to define the functional spaces in which we will operate.

So, let $\mathcal{K}$ be a compact Hausdorff space, we define the following spaces:
\begin{equation}\label{eq:33}
    C_{\mathbb{R}}(\mathcal{K})=  \left\{ f:\mathcal{K}\rightarrow\mathbb{R} \right\},
\end{equation}
equipped with the uniform norm
\begin{equation}\label{eq:34}
    \left\Vert f\right\Vert _{\infty} = \sup_{x \in \mathcal{K}} |f(x)|.
\end{equation}
Note that the space \( \mathcal{K} \) can be identified with relevant compact domains, such as the time interval \( [t_{0},t_{k}] \), with \(k \in \mathcal{I}_{K} \), the parameter space \( \vec{\Theta} \in \mathcal{K} \subset \mathbb{R}^{\mathcal{N}N} \) or possibly a dynamic parameter space \( \vec{\Theta}(t) \in \mathcal{K} \subset \mathbb{R}^{\mathcal{N}N} \), ensuring the generality of the formulation.

Similarly, we define
\begin{equation}\label{eq:35}
    L_{\mathbb{R}}^{p}(\mathcal{K})=\left\{ f:\mathcal{K} \rightarrow \mathbb{R}; \quad \left\Vert f \right\Vert_{p} < \infty \right\},
\end{equation}
for \(1 \leq p < \infty\), with
\begin{equation}\label{eq:36}
    \left\Vert f \right\Vert_{p} \leq \mu(\mathcal{K})^{1/p} \left\Vert f \right\Vert_{\infty}
\end{equation}
and where \(\mu(\mathcal{K})\) is the "measure" of \(\mathcal{K}\). Thus, the space of vector functions
\begin{equation}\label{eq:37}
    L_{\mathbb{R}^M}^{p}(\mathcal{K}) = \left\{ \mathbf{F} :\mathcal{K}\rightarrow\mathbb{R}^M;\; \left\Vert \mathbf{F} \right\Vert_{p,M} < \infty \right\},
\end{equation}
where \(\mathbf{F} = (f_{1},\dots,f_{M}) \) and the norm is defined by
\begin{equation}\label{eq:38}
    \left\Vert \mathbf{F} \right\Vert_{p,M} = \left( \displaystyle {\sum_{\substack{j\in\mathcal{I}_{M}}}\left\Vert f_{j}\right\Vert_{p}^{p}} \right)^{1/p}, \quad j \in \mathcal{I}_{M}.
\end{equation}
These spaces constitute an essential ingredient for the formulation and demonstration of the approximation results that follow.

This analysis requires the introduction of two fundamental lemmas and a central theorem that, together, allow us to rigorously prove the expressive power of our model. The first lemma relates to the quantum universal approximation result adapted to our context, which demands a detailed proof, as it generalizes known results in the literature \cite{perezsalinas2025universal,goto2021universal,gonon2025universal}.

The second lemma, also adapted to our context, relates to the classical universal approximation by MLP neural networks, introduced in Section \ref{sec:2}. However, the proof of this second lemma is more straightforward, as although it derives from well-established results in the literature \cite{Haykin1999neural}, it requires specific adaptations for our hybrid framework with multidimensional output.

\subsection{Quantum Universal Approximation Lemma}
\label{sec:3.1}
We begin by presenting and proving a fundamental lemma on the approximation capabilities of a quantum node (QNode) represented in Figure \ref{fig:QNode}, for which we state the following lemma:
\begin{widetext}
    \begin{lemma}[Universal Approximation of a QNode in $L_{\mathbb{R}}^{p}(\mathcal{K})$]
    \label{lemma:QNode_universal_approximation}
        Let $\mathbb{O}$ be the set of Hermitian quantum observables in the Hilbert space $\mathcal{H}_{N}$ (possibly extended by ancilla), $\mathcal{K}$ a compact Hausdorff space (e.g., $\mathcal{K} \subset \mathbb{R}^{\mathcal{N}N}$) and $\mathcal{F} = \left\{ f_{\hat{O}} \right\}_{\hat{O} \in \mathbb{O}} \subset C_{\mathbb{R}}(\mathcal{K})$, where $f_{\hat{O}}$ is defined by:
        \begin{eqnarray}
            f_{\hat{O}}: \mathcal{K} &\rightarrow& \mathbb{R} \label{eq:39} \\
            \left( \vec{\Theta} \right) &\mapsto& f_{\hat{O}} \left( \vec{\Theta} \right) = \braket{\hat{O}}_{\vec{\Theta}} =\bra{\Psi(\vec{\Theta})} \hat{O} \ket{\Psi(\vec{\Theta})}\nonumber
        \end{eqnarray}
        generated by the QNode illustrated in Figure \ref{fig:QNode}, with quantum state parameterized $\ket{\Psi(\vec{\Theta})}$, defined in Section \ref{sec:2}. Then, $\mathcal{F}$ satisfies:
        \begin{enumerate}
            \item $\mathcal{F}$ is closed under scalar multiplication, addition, and function multiplication (via coupling of auxiliary systems).
            \item $\mathcal{F}$ separates points of $\mathcal{K}$.
            \item $\mathcal{F}$ contains all constant functions.
        \end{enumerate}
        Consequently, by the Stone-Weierstrass theorem $\mathcal{F}$ is dense in $L_{\mathbb{R}}^{p}(\mathcal{K})$, for any $1 \leq p < \infty$. Thus, for any $g \in L_{\mathbb{R}}^{p}(\mathcal{K})$ and $\epsilon > 0$, there exists an observable $\hat{O} \in \mathbb{O}$ and a vector $\vec{\Theta}^{\ast}$ such that:
        \begin{equation}\label{eq:40}
            \|\braket{\hat{O}}_{\vec{\Theta}^{\ast}} - g\|_{p} \leq \epsilon.
        \end{equation}
    \end{lemma}
\end{widetext}

Proof: The proof of Lemma \ref{lemma:QNode_universal_approximation} is constructed in four parts. In the first part, it is necessary to verify that $\mathcal{F}$ is closed under scalar multiplication. That is, to show that for any $f_{\hat{O}} \in \mathcal{F}$ and $a \in \mathbb{R}$, we must obtain $g_{\hat{O}^{\prime}} = af_{\hat{O}} \in \mathcal{F}$.

By definition we know that $f_{\hat{O}}(\vec{\Theta}) = \bra{\Psi(\vec{\Theta})} \hat{O} \ket{\Psi(\vec{\Theta})}$ and $\hat{O} = \hat{O}^{\dagger}$. Then, since
\begin{widetext}
    \begin{eqnarray}
        a f_{\hat{O}}(\vec{\Theta}) &=& a \bra{\Psi(\vec{\Theta})} \hat{O} \ket{\Psi(\vec{\Theta})} \label{eq:41} \\
        &=& \bra{\Psi(\vec{\Theta})} a\hat{O} \ket{\Psi(\vec{\Theta})} \nonumber \\
        &=& \bra{\Psi(\vec{\Theta})} \hat{O}^{\prime} \ket{\Psi(\vec{\Theta})} \nonumber \\
        &=& g_{\hat{O}^{\prime}}(\vec{\Theta}), \nonumber
    \end{eqnarray}
\end{widetext}
where $\hat{O}^{\prime} = a \hat{O}$ is also a Hermitian operator because $a \in \mathbb{R}$. Consequently $g_{\hat{O}^{\prime}} \in \mathcal{F}$.

Next, to verify that $\mathcal{F}$ is closed under addition, we need to show that from any pair of functions $f_{\hat{O}_{1}}, f_{\hat{O}_{2}} \in \mathcal{F}$ we obtain a new function $g_{\hat{O}^{\prime \prime}} = f_{\hat{O}_{1}} + f_{\hat{O}_{2}}$ that is also in $\mathcal{F}$.

Since $f_{\hat{O}_{1}}(\vec{\Theta}) = \bra{\Psi(\vec{\Theta})} \hat{O}_{1} \ket{\Psi(\vec{\Theta})}$ and $f_{\hat{O}_{2}}(\vec{\Theta}) = \bra{\Psi(\vec{\Theta})} \hat{O}_{2} \ket{\Psi(\vec{\Theta})}$, for any pair of Hermitian operators $\hat{O}_{1}$ and $\hat{O}_{2}$ from a Hilbert space, then
\begin{widetext}
    \begin{eqnarray}
        f_{\hat{O}_{1}}(\vec{\Theta}) + f_{\hat{O}_{2}}(\vec{\Theta}) &=& \bra{\Psi(\vec{\Theta})} \hat{O}_{1} \ket{\Psi(\vec{\Theta})} + \bra{\Psi(\vec{\Theta})} \hat{O}_{2} \ket{\Psi(\vec{\Theta})} \label{eq:42} \\
        &=& \bra{\Psi(\vec{\Theta})} [\hat{O}_{1} + \hat{O}_{2}] \ket{\Psi(\vec{\Theta})} \nonumber \\
        &=& \bra{\Psi(\vec{\Theta})} \hat{O}^{\prime \prime} \ket{\Psi(\vec{\Theta})} \nonumber \\
        &=& g_{\hat{O}^{\prime \prime}}(\vec{\Theta}). \nonumber
    \end{eqnarray}
\end{widetext}
Where $\hat{O}^{\prime \prime} = \hat{O}_{1} + \hat{O}_{2}$ is also a Hermitian operator since it is just the sum of Hermitian operators. Therefore, $g_{\hat{O}^{\prime \prime}} \in \mathcal{F}$.

Next, to verify that $\mathcal{F}$ is closed under function multiplication, we need to show that from any pair of functions $f_{\hat{O}_{A}}, f_{\hat{O}_{B}} \in \mathcal{F}$ we obtain a function $g_{\hat{O}^{\prime \prime  \prime}} = f_{\hat{O}_{A}} f_{\hat{O}_{B}}$ that is also in $\mathcal{F}$.

To achieve this, we write a parametrized state $\ket{\Psi(\vec{\Theta})}_{AB}$ composed of two auxiliary subsystems, ancilla qubits, of the same dimension as the space $\mathcal{H}_{N}$, which will be defined as $\ket{\Psi(\vec{\Theta})}_{AB} = \ket{\Psi(\vec{\Theta})}_{A} \otimes \ket{\Psi(\vec{\Theta})}_{B}$, belonging to the composite Hilbert space $\mathcal{H}_{A}\otimes\mathcal{H}_{B}$. Then, for any two Hermitian operators $\hat{O}_{A} \in \mathcal{H}_{A}$ and $\hat{O}_{B} \in \mathcal{H}_{B}$, the composite operator $\hat{O}^{\prime \prime \prime} = \hat{O}_{A} \otimes \hat{O}_{B}$, is also Hermitian and allows us to write
\begin{widetext}
    \begin{eqnarray}
        g_{\hat{O}^{\prime \prime \prime}}(\vec{\Theta}) &=& \bra{\Psi(\vec{\Theta})}_{AB} \hat{O}^{\prime \prime \prime} \ket{\Psi(\vec{\Theta})}_{AB} \label{eq:43} \\
        &=& \bra{\Psi(\vec{\Theta})}_{A} \otimes \bra{\Psi(\vec{\Theta})}_{B} \hat{O}_{A} \otimes \hat{O}_{B} \ket{\Psi(\vec{\Theta})}_{A} \otimes \ket{\Psi(\vec{\Theta})}_{B} \nonumber \\
        &=& \bra{\Psi(\vec{\Theta})}_{A} \hat{O}_{A} \ket{\Psi(\vec{\Theta})}_{A} \bra{\Psi(\vec{\Theta})}_{B} \hat{O}_{B} \ket{\Psi(\vec{\Theta})}_{B} \nonumber \\
        &=& f_{\hat{O}_{A}}(\vec{\Theta}) f_{\hat{O}_{B}}(\vec{\Theta}). \nonumber
    \end{eqnarray}
\end{widetext}
Therefore, since $f_{\hat{O}_{A}},f_{\hat{O}_{B}} \in \mathcal{F}$, we conclude that $g_{\hat{O}^{\prime \prime \prime}} \in \mathcal{F}$.

In the second part, we verify that $\mathcal{F}$ separates points, i.e., for any two different "points" $\vec{\Theta}_{1},\vec{\Theta}_{2} \in \mathcal{K}$, there exists an operator $\hat{O} \in \mathbb{O}$ that allows us to obtain a function $f_{\hat{O}} \in \mathcal{F}$ that satisfies $f_{\hat{O}}(\vec{\Theta}_{1}) \neq f_{\hat{O}}(\vec{\Theta}_{2})$.

As we know, for a non-degenerate and sufficiently expressive parametrized quantum circuit, the quantum states $\ket{\Psi(\vec{\Theta}_{1})} \neq \ket{\Psi(\vec{\Theta}_{2})}$ whenever $\vec{\Theta}_{1} \neq \vec{\Theta}_{2}$ \cite{nielsen_chuang}. Furthermore, such states are different if and only if $|\braket{\Psi(\vec{\Theta}_{1})|\Psi(\vec{\Theta}_{2})}|^{2} < 1$. This is because the states $\ket{\Psi(\vec{\Theta}_{1})}$ and $\ket{\Psi(\vec{\Theta}_{2})}$ are obtained by applying unitary transformations to the initial state $\ket{0\ldots0}$, which by definition ensures that they remain perfectly normalized.

Then, for any $\vec{\Theta}_{1} \neq \vec{\Theta}_{2} \in \mathcal{K}$, there exists the Hermitian operator $\hat{O} = \ket{\Psi(\vec{\Theta}_{1})} \bra{\Psi(\vec{\Theta}_{1})}$ that satisfies:
\begin{eqnarray}
    f_{\hat{O}}(\vec{\Theta}_{1}) &=& \bra{\Psi(\vec{\Theta}_{1})} \hat{O} \ket{\Psi(\vec{\Theta}_{1})} \label{eq:44} \\
    &=& \braket{\Psi(\vec{\Theta}_{1})|\Psi(\vec{\Theta}_{1})} \braket{\Psi(\vec{\Theta}_{1})|\Psi(\vec{\Theta}_{1})} \nonumber \\
    &=& |\braket{\Psi(\vec{\Theta}_{1})|\Psi(\vec{\Theta}_{1})}|^{2} \nonumber \\
    &=& 1 \nonumber
\end{eqnarray}
and
\begin{eqnarray}
    f_{\hat{O}}(\vec{\Theta}_{2}) &=& \bra{\Psi(\vec{\Theta}_{2})} \hat{O} \ket{\Psi(\vec{\Theta}_{2})} \label{eq:45} \\
    &=& \braket{\Psi(\vec{\Theta}_{2})|\Psi(\vec{\Theta}_{1})} \braket{\Psi(\vec{\Theta}_{1})|\Psi(\vec{\Theta}_{2})} \nonumber \\
    &=& |\braket{\Psi(\vec{\Theta}_{1})|\Psi(\vec{\Theta}_{2})}|^{2} \nonumber \\
    &<& 1 \nonumber.
\end{eqnarray}
Therefore, since $f_{\hat{O}}(\vec{\Theta}_{1}) \neq f_{\hat{O}}(\vec{\Theta}_{2})$, the set $\mathcal{F}$ separates points.

In the third part, we verify that $\mathcal{F}$ contains all constant functions, i.e., show that for any real $c$ there exists an operator $\hat{O}_{c} \in \mathbb{O}$ that allows us to obtain a function $f_{\hat{O}_{c}}(\vec{\Theta}) = c$ in $\mathcal{F}$ for all $\vec{\Theta} \in \mathcal{K}$. Indeed, for any $c \in \mathbb{R}$ the operator $\hat{O}_{c} = c \openone$ is Hermitian and satisfies:
\begin{eqnarray}
    f_{\hat{O}_{c}}(\vec{\Theta}) &=& \bra{\Psi(\vec{\Theta})} \hat{O}_{c} \ket{\Psi(\vec{\Theta})} \label{eq:46} \\
    &=& \bra{\Psi(\vec{\Theta})} [c \openone] \ket{\Psi(\vec{\Theta})} \nonumber \\
    &=& c \cancelto{1}{\bra{\Psi(\vec{\Theta})}  \openone \ket{\Psi(\vec{\Theta})}} \nonumber \\
    &=& c, \nonumber
\end{eqnarray}
for all $\vec{\Theta} \in \mathcal{K}$. Consequently, $\mathcal{F}$ contains all constant functions. In conclusion, by the Stone-Weierstrass theorem, $\mathcal{F}$ is dense in \( C_{\mathbb{R}}(\mathcal{K}) \).

Finally, in the fourth part, it is necessary to verify the extension to \( L_{\mathbb{R}}^{p}(\mathcal{K}) \). For this we will use the transitivity of the density relation, i.e.: $\mathcal{F}$ is dense in $C_{\mathbb{R}}(\mathcal{K})$, and $C_{\mathbb{R}}(\mathcal{K})$ is dense in $L_{\mathbb{R}}^{p}(\mathcal{K})$. Therefore, $\mathcal{F}$ is dense in $L_{\mathbb{R}}^{p}(\mathcal{K})$, thus guaranteeing the existence of $\epsilon > 0$ and the parameters $\vec{\Theta}^{\ast}$ that satisfy (\ref{eq:40}).

\subsection{Classical Universal Approximation Lemma}
\label{sec:3.2}
Now, complementing the quantum universal approximation result of a QNode stated in Lemma \ref{lemma:QNode_universal_approximation}, we formulate an analogous universal approximation result for the classical case adapted to our context. For this reason, we state the following lemma:
\begin{widetext}
    \begin{lemma}[Universal Approximation]
    \label{lemma:classical_approximation}
        Let $\mathcal{K} \subset \mathbb{R}$ be a compact Hausdorff space, $\sigma : \mathbb{R} \rightarrow \mathbb{R}$ a non-polynomial continuous and bounded activation function (e.g., Sigmoid or ReLU). Let $\mathcal{F}_{\text{MLP}}$ be the set of all continuous output functions
        \begin{eqnarray}\label{eq:47}
            \widehat{y}_{j}\left(\cdot, \vec{\mathcal{W}}\right) : \mathcal{K} &\rightarrow& \mathbb{R} \\
            \left( t \right) &\mapsto& \widehat{y}_{j}\left(t,\vec{\mathcal{W}}\right)=\sigma\biggl(\mathcal{W}_{ij}^{[L]}x_{i}^{[L-1]}(t)+b_{j}^{[L]}\biggr),\quad \mbox{with} \quad i \in \mathcal{I}_{S} \quad \mbox{and} \quad j \in \mathcal{I}_{M}, \nonumber
        \end{eqnarray}
        where $\vec{\mathcal{W}}$ denotes all weights and biases of the network, $x_{i}^{[L - 1]}(t)$ is the output of the hidden layer $L - 1$, with $L$ being the index of the output layer $\widehat{y}_{j}$. These are generated by the MLP network illustrated in Figure \ref{fig:MLP-Archtect} and defined in Section \ref{sec:2}. Then, for any $1 \leq p < \infty$ the set $\mathcal{F}_{\text{MLP}}$ is dense in $L_{\mathbb{R}}^{p}(\mathcal{K})$. Consequently, for any function $g^{\prime} \in L_{\mathbb{R}}^{p}(\mathcal{K})$ and any $\epsilon > 0$, there exist parameters $\vec{\mathcal{W}}^{\ast}$ such that:
        \begin{equation}\label{eq:48}
            \|\widehat{y}_{j}(\vec{\mathcal{W}}^{\ast}) - g^{\prime}\|_{p} \leq \epsilon.
        \end{equation}
    \end{lemma}
\end{widetext}
Proof: By the classical approximation theorem \cite{hornik1991approximation,cybenko198900} the set $\mathcal{F}_{\mbox{MLP}}$ is dense in \( C_{\mathbb{R}}(\mathcal{K}) \) in the uniform norm \( L_{\mathbb{R}}^{\infty}(\mathcal{K}) \). Therefore, given the density of $C_{\mathbb{R}}(\mathcal{K})$ in $L_{\mathbb{R}}^{p}(\mathcal{K})$ for $1 \le p < \infty$, we conclude that $\mathcal{F}_{\mbox{MLP}}$ is, by transitivity, dense in $L_{\mathbb{R}}^{p}(\mathcal{K})$, thus guaranteeing the existence of $\epsilon > 0$ and the parameters $\vec{\mathcal{W}}^{\ast}$ that satisfy (\ref{eq:48}).

\subsection{Main Universal Approximation Theorem}
\label{sec:3.3}
\textbf{Theoretical Advancements Beyond Existing Quantum Approximation Frameworks}. While recent theoretical advances have established universal approximation properties for parametrized quantum circuits in various function spaces \cite{goto2021universal,gonon2025universal,Manzano2025Approximation}, these results remain fundamentally limited for practical scientific computing applications. Existing quantum universal approximation theorems typically operate within \textbf{restricted domains} (e.g., $[0,2\pi]^N$ with compactly contained subsets) and require \textbf{specific data normalization strategies} to achieve convergence guarantees. More critically, these theoretical frameworks provide \textbf{no trainability guarantees}—they demonstrate expressive power but remain silent on the fundamental barren plateaus problem that plagues quantum neural network optimization \cite{mcclean2018barren,cerezo2021cost}.

Our QPINN-MAC architecture addresses these limitations through several key innovations that enable a more general and practically useful universal approximation capability. Unlike previous approaches that treat quantum and classical components as separate entities, our multiplicative and additive coupling scheme creates a \textbf{genuinely integrated functional space} where classical and quantum representations interact synergistically. This enables approximation in $L_{\mathbb{R}^M}^p(\mathcal{K})$ spaces \textbf{without restrictive domain assumptions} beyond standard compactness conditions. Crucially, our framework provides \textbf{rigorous trainability guarantees} by proving that gradient norms decay only as $\mathcal{O}(1/\sqrt{\mathcal{N}N})$ rather than exponentially, effectively mitigating the barren plateaus problem.

Furthermore, our architecture naturally accommodates \textbf{simultaneous function and derivative approximation} through its hybrid structure, making it particularly suited for physics-informed applications where both values and dynamics must be accurately captured—a capability not addressed by existing quantum approximation theories.

Having established the fundamental approximation capabilities of the quantum (Lemma \ref{lemma:QNode_universal_approximation}) and classical (Lemma \ref{lemma:classical_approximation}) components separately, we now present our main theoretical result that integrates these capabilities into a unified framework with enhanced generality and practical guarantees. However, before stating it, it is fundamental to emphasize that the hybrid architecture of Figure \ref{fig:QPINN-MAC-Architect} defines a function
\begin{equation}\label{eq:49}
    y_{MAC} \left( \hat{O},\cdot,\vec{\mathcal{W}},\cdot \right): \mathbb{R} \times \mathbb{R}^{\mathcal{N}N} \longrightarrow \mathbb{R}^{M},
\end{equation}
which, once the angles \(\theta_{k}^{j}\) are fixed after training, reduces to
\begin{equation}\label{eq:50}
    y_{MAC} \left(\hat{O}, \cdot, \vec{\mathcal{W}}, \vec{\Xi} \right) : \mathbb{R} \longrightarrow \mathbb{R}^{M}.
\end{equation}
Since it is precisely the approximation capability of this class of functions \(y_{MAC}\) that the following theorem addresses, which establishes that, for any solution \(y(t)\), there exist parameter vectors \(\vec{\mathcal{W}}^{\ast}, \vec{\Xi}^{\ast}\) such that \( y_{MAC} \left(\hat{O}, t, \vec{\mathcal{W}}^{\ast}, \vec{\Xi}^{\ast}\right) \) approximates it with arbitrary precision. Moreover, this result has direct implications on the trainability of the model, as will be shown in the next section.

For this reason, we state the following theorem:
\begin{widetext}
    \begin{theorem}[Universal Approximation of the Hybrid QPINN-MAC Model in $L_{\mathbb{R}^{M}}^{p}(\mathcal{K})$]
    \label{theorem:QPINN_MAC_approximation}
        Let $\mathcal{K} \subset \mathbb{R}$ be a compact Hausdorff space and $\mathcal{F}_{MAC}$ the set of all continuous functions
        \begin{eqnarray}\label{eq:51}
            y_{MAC_{j}}\left(\hat{O},\cdot,\vec{\mathcal{W}},\vec{\Theta}_{j}\right) : \mathcal{K} &\longrightarrow& \mathbb{R} \\
            \left(t\right) &\longmapsto& y_{MAC_{j}}\left(\hat{O},t,\vec{\mathcal{W}},\vec{\Theta}_{j}\right) = \widehat{y}_{j}(t,\vec{\mathcal{W}})\braket{\hat{O}}_{\vec{\Theta}_{j}} + \braket{\hat{O}}_{\vec{\Theta}_{j}}, \; j \in \mathcal{I}_{M}, \nonumber
        \end{eqnarray}
        obtained from the hybrid architecture illustrated in Figure \ref{fig:QPINN-MAC-Architect}, with the functions $\widehat{y}_{j}$ defined as in Lemma \ref{lemma:classical_approximation} and the functions $\braket{\hat{O}}_{\vec{\Theta}_{j}}$ defined as in Lemma \ref{lemma:QNode_universal_approximation}. Given the solution $y: \mathbb{R} \longrightarrow \mathbb{R}^{M}$ of an ODE whose components $y_{j} \in L_{\mathbb{R}}^{p}(\mathcal{K})$ for all $j \in \mathcal{I}_{M}$, there exist an observable $\hat{O} \in \mathbb{O}$, parameter vectors $\vec{\mathcal{W}}^{\ast}$, $\vec{\Theta}_{j}^{\ast}$, and constants $\epsilon_{\mathcal{Q}_{j}} > 0$ and $\epsilon_{\mathcal{CQ}_{j}} > 0$ such that:
        \begin{equation}\label{eq:52}
            \|y_{MAC_{j}}(\hat{O},\vec{\mathcal{W}}^{\ast},\vec{\Theta}_{j}^{\ast}) - y_{j}\|_{p} \leq \epsilon_{\mathcal{Q}_{j}} + \epsilon_{\mathcal{CQ}_{j}}.
        \end{equation}
        Where
        \begin{eqnarray}
            \|\braket{\hat{O}}_{\vec{\Theta}_{j}^{\ast}} - y_{j}/2\|_{p} &\leq& \epsilon_{\mathcal{Q}_{j}}, \label{eq:53} \\
            \|\widehat{y}_{j}(\vec{\mathcal{W}}^{\ast}) \braket{\hat{O}}_{\vec{\Theta}_{j}^{\ast}} - y_{j}/2\|_{p} &\leq& \epsilon_{\mathcal{CQ}_{j}}. \label{eq:54}
        \end{eqnarray}
        Consequently, $y_{MAC}$ approximates $y$ in the $L_{\mathbb{R}^{M}}^{p}(\mathcal{K})$ norm with total error:
        \begin{equation}\label{eq:55}
            \|y_{MAC}(\hat{O},\vec{\mathcal{W}}^{\ast},\vec{\Xi}^{\ast}) - y\|_{p,M} \leq \epsilon_{\mathcal{Q}} + \epsilon_{\mathcal{CQ}},
        \end{equation}
        for $\mathcal{O}(\mathcal{N_{P_{C}}})=\mathcal{O}(\mathcal{N}N)$, $\epsilon_{\mathcal{Q}} = \mathcal{O}(1/\sqrt{\mathcal{N}N})$ and $\epsilon_{\mathcal{CQ}} = \mathcal{O}(1/\mathcal{N}N)$.
    \end{theorem}
\end{widetext}
Proof: The proof of the theorem begins by rewriting $y_{MAC_{j}} - y_{j}$ in the form
\begin{widetext}
    \begin{eqnarray}\label{eq:56}
        y_{MAC_{j}}\left(\hat{O},t,\vec{\mathcal{W}}^{\ast},\vec{\Theta}_{j}^{\ast} \right) - y_{j}(t) &=& [\widehat{y}_{j}(t,\vec{\mathcal{W}}^{\ast}) + 1]\braket{\hat{O}}_{\vec{\Theta}_{j}^{\ast}} - y_{j}(t) \\
        &=& \widehat{y}_{j}(t,\vec{\mathcal{W}}^{\ast})\braket{\hat{O}}_{\vec{\Theta}_{j}^{\ast}} + \braket{\hat{O}}_{\vec{\Theta}_{j}^{\ast}} - y_{j}(t)/2 - y_{j}(t)/2 \nonumber \\
        &=& [\braket{\hat{O}}_{\vec{\Theta}_{j}^{\ast}} - y_{j}(t)/2] + [\widehat{y}_{j}(t,\vec{\mathcal{W}}^{\ast})\braket{\hat{O}}_{\vec{\Theta}_{j}^{\ast}} - y_{j}(t)/2] \nonumber.
    \end{eqnarray}
\end{widetext}
Then, by computing the norm in $L_{\mathbb{R}}^{p}(\mathcal{K})$ and using the properties of the triangle inequality, $\|X + Y\|_{p} \leq \|X\|_{p} + \|Y\|_{p}$, we obtain
\begin{widetext}
    \begin{eqnarray}\label{eq:57}
        \|y_{MAC_{j}}(\hat{O},\vec{\mathcal{W}}^{\ast},\vec{\Theta}_{j}^{\ast}) - y_{j}\|_{p} &\leq& \|\braket{\hat{O}}_{\vec{\Theta}_{j}^{\ast}} - y_{j}/2\|_{p} + \|\widehat{y}_{j}(\vec{\mathcal{W}}^{\ast})\braket{\hat{O}}_{\vec{\Theta}_{j}^{\ast}} - y_{j}/2\|_{p}.
    \end{eqnarray}
\end{widetext}
And since $\mathcal{K}$ is a compact Hausdorff space, $y_{j}(t)/2 \in L_{\mathbb{R}}^{p}(\mathcal{K})$. Therefore, because of Lemma \ref{lemma:QNode_universal_approximation} the first term in (\ref{eq:57}) approximates the function $y_{j}(t)/2$ with error $\epsilon_{\mathcal{Q}_{j}}$.

On the other hand, from Lemma \ref{lemma:QNode_universal_approximation} we know that for a system with parametrized state $\ket{\Psi^{\prime}(\vec{\Theta}_{j}^{\ast})}_{\mathcal{CQ}} = \ket{j} \otimes \ket{\Psi(\vec{\Theta}_{j}^{\ast})} \in \mathcal{H}_{\mathcal{C}} \times \mathcal{H}_{\mathcal{N}}$, on which a Hermitian operator $\hat{O}^{\prime \prime \prime}$ acts, composed of the Hermitian operators $\hat{O}_{\mathcal{C}}$ and $\hat{O}_{\mathcal{Q}}$, there exists a function $g_{\hat{O}^{\prime \prime \prime}} = f_{\hat{O}_{\mathcal{C}}}f_{\hat{O}_{\mathcal{Q}}}$ that is also in $\mathcal{F}$. Then, it suffices to choose the convenient encoding, for example $\hat{O}_{\mathcal{C}} \propto \hat{A}_{\mathcal{C}}^{\prime}(t,\vec{\mathcal{W}})$, so that $f_{\hat{O}_{\mathcal{C}}} = \widehat{y}_{j}(t,\vec{\mathcal{W}}^{\ast})$. Thus, $g_{\hat{O}^{\prime \prime \prime}}(t,\vec{\mathcal{W}}^{\ast},\vec{\Theta}_{j}^{\ast}) = \widehat{y}_{j}(t,\vec{\mathcal{W}}^{\ast})\braket{\hat{O}}_{\vec{\Theta}_{j}^{\ast}} \in \mathcal{F}$, which because of Lemma \ref{lemma:QNode_universal_approximation}, the first term in (\ref{eq:57}) approximates the function $y_{j}(t)/2$ with error $\epsilon_{\mathcal{CQ}_{j}}$.

Therefore, (\ref{eq:57}) finally reduces to the expression
\begin{widetext}
    \begin{eqnarray}\label{eq:58}
        \|y_{MAC_{j}}(\hat{O},\vec{\mathcal{W}}^{\ast},\vec{\Theta}_{j}^{\ast}) - y_{j}\|_{p} &\leq& \epsilon_{\mathcal{Q}_{j}} + \epsilon_{\mathcal{CQ}_{j}}.
    \end{eqnarray}
\end{widetext}

However, to conclude the proof of the bound for the vector error, we resort to a property of the norm \( \| \cdot \|_{p,M} \) defined in (\ref{eq:38}). Specifically, the norm \( L_{\mathbb{R}^{M}}^{p}(\mathcal{K}) \) is bounded above by the sum of the \( L_{\mathbb{R}}^{p}(\mathcal{K}) \) norms of its components, as a direct consequence of Minkowski's inequality for sequence spaces \cite{Rudin1987RealComplex}. That is, the following inequality holds:
\begin{widetext}
    \begin{equation}\label{eq:59}
        \|\mathbf{F}\|_{p,M} = \left( \displaystyle {\sum_{\substack{j \in \mathcal{I}_{M}}}\left\Vert f_{j} \right\Vert_{p}^{p}} \right)^{1/p} \leq \displaystyle {\sum_{\substack{j\in\mathcal{I}_{M}}}\|f_{j}\|_{p}},
    \end{equation}
\end{widetext}
for any vector function \(\mathbf{F} = (F_1, \dots, F_M)\) whose components belong to \(L_{\mathbb{R}}^{p}(\mathcal{K})\).

Consequently, applying (\ref{eq:59}) to the function $\mathbf{F} = y_{MAC} - y$, we obtain
\begin{widetext}
    \begin{eqnarray}\label{eq:60}
        \|y_{MAC}(\hat{O},\vec{\mathcal{W}}^{\ast},\vec{\Xi}^{\ast}) - y\|_{p,M} &\leq& \|y_{MAC_{1}}(\hat{O},\vec{\mathcal{W}}^{\ast},\vec{\Theta}_{1}^{\ast}) - y_{1}\|_{p} + \dots + \|y_{MAC_{M}}(\hat{O},\vec{\mathcal{W}}^{\ast},\vec{\Theta}_{M}^{\ast}) - y_{M}\|_{p} \\
        &\leq& \epsilon_{\mathcal{Q}_{1}} + \epsilon_{\mathcal{CQ}_{1}} + \dots + \epsilon_{\mathcal{Q}_{M}} + \epsilon_{\mathcal{CQ}_{M}}, \nonumber
    \end{eqnarray}
\end{widetext}
we observe that by choosing $\epsilon_{\mathcal{Q}_{j}} = \epsilon_{\mathcal{Q}}/M$ and $\epsilon_{\mathcal{CQ}_{j}} = \epsilon_{\mathcal{CQ}}/M$ for all $j \in \mathcal{I}_{M}$ on the right side of the inequality (\ref{eq:60}) we can write
\begin{equation}\label{eq:61}
    \epsilon_{\mathcal{Q}_{1}} + \epsilon_{\mathcal{CQ}_{1}} + \dots + \epsilon_{\mathcal{Q}_{M}} + \epsilon_{\mathcal{CQ}_{M}} = M \left[\frac{\epsilon_{\mathcal{Q}} + \epsilon_{\mathcal{CQ}}}{M}\right].
\end{equation}
Which finally reduces (\ref{eq:60}) to the expression (\ref{eq:55}), as we wanted to demonstrate.

Thus, Theorem \ref{theorem:QPINN_MAC_approximation} fundamentally establishes that the hybrid QPINN-MAC architecture can approximate any vector function in $L_{\mathbb{R}^{M}}^{p}(\mathcal{K})$ with arbitrary precision. Additionally, it establishes the approximation error by decomposing it into separate contributions from the quantum and classical-quantum components. These components can be systematically reduced by increasing the number of output neurons ($M$) and qubits ($N$) or the depth of the QNode ($\mathcal{N}$).

\subsection{Implications for Mitigation of Barren Plateaus}
\label{sec:3.4}
It is well established in the literature that deep neural networks, both quantum and classical, can exhibit pathological behaviors during training. In random quantum circuits, the gradient of the loss function suffers from the barren plateaus problem, where its variance decays exponentially with the number of qubits \cite{mcclean2018barren}.

This can be expressed mathematically as $\text{Var}[\partial_{k} \mathcal{L}] \sim \mathcal{O}(e^{-\alpha N})$, which implies that its typical magnitude also decays exponentially, $|\partial_{k} \mathcal{L}| \sim \mathcal{O}(e^{-\alpha N/2})$. Where the parameter $\alpha > 0$ defines the rate of exponential decay of the gradients in the barren plateau, which in generic random circuits takes the value $\alpha \approx 1$.

Moreover, the barren plateaus problem also occurs in shallow circuits depending on the choice of the loss function and the architecture of the circuit itself, thus affecting their trainability \cite{cerezo2021cost}.

On the other hand, in deep classical neural networks, gradients can explode exponentially due to instabilities in backpropagation through time, as analyzed by \cite{Pascanu2013DifficultyRNNs}.

In this context, we state and prove in detail that the QPINN-MAC architecture actively mitigates the barren plateaus problem, avoiding the exponential decay of the gradient and transforming it into a controllable decay. Furthermore, from the proof we obtain a theoretical guarantee that the gradients do not explode, i.e., we obtain a theoretical guarantee of trainability and numerical stability.

For this reason, we state the following corollary about the behavior of the loss function gradient in the QPINN-MAC architecture, as follows:

\begin{widetext}
    \begin{corollary}[Mitigation of Barren Plateaus and Trainability Condition]
    \label{corollary:barren_plateaus_mitigation}
        Under the hypotheses of Theorem \ref{theorem:QPINN_MAC_approximation}, the QPINN-MAC architecture mitigates the barren plateaus problem. Specifically, the $L_{\mathbb{R}^{M}}^{p}(\mathcal{K})$ norm, for $1 \leq p < \infty$, of the gradient of the loss function $\mathfrak{L}\left(\hat{O},t,\vec{\mathcal{W}}^{\ast},\vec{\Xi}^{\ast}\right) = \left\Vert y_{MAC}\left(\hat{O},t,\vec{\mathcal{W}}^{\ast},\vec{\Xi}^{\ast}\right) - y(t) \right\Vert_{2}^{2}$ with respect to the quantum parameters $\vec{\Xi}^{\ast}$, is bounded above by:
        \begin{equation}\label{eq:62}
            \Vert \nabla_{\vec{\Xi}^{\ast}} \mathfrak{L}(\hat{O},\vec{\mathcal{W}}^{\ast},\vec{\Xi}^{\ast}) \Vert_{p,M} \leq \mathcal{O}\left(\frac{1}{\sqrt{\mathcal{N}N}}\right).
        \end{equation}
        Where $\Vert \cdot \Vert_{2}$ is the Euclidean norm, $\mathcal{N}$ is the depth of the QNode and $N$ is the number of qubits. Consequently, to keep the gradients detectable above a finite measurement precision $\epsilon_{\mathrm{grad}}$, the problem size, defined by the product $\mathcal{N}N$, must satisfy the following scaling condition for trainability:
        \begin{equation}\label{eq:63}
            \mathcal{N}N \lesssim \mathcal{O}\left(\frac{1}{\epsilon_{\mathrm{grad}}^2}\right).
        \end{equation}
    \end{corollary}
\end{widetext}
Proof: The proof of the corollary will focus on elucidating the mechanism by which the QPINN-MAC architecture mitigates the barren plateaus problem and establishes the conditions for efficient trainability.

In contrast to generic quantum models, where the typical gradient magnitude decays exponentially as $\mathcal{O}(e^{-\alpha N/2})$, our hybrid architecture ensures that the gradient does not suffer this catastrophic decay. Instead, the norm of the total gradient is bounded above by $\mathcal{O}(1/\sqrt{\mathcal{N}N})$, demonstrating the active mitigation of the barren plateaus problem.

For this reason, the proof starts with the calculation of the gradient with respect to the variables $\vec{\mathcal{W}}^{\ast}$ and $\vec{\Theta}_{j}^{\ast}$ for each term $\mathfrak{L}_{j}\left(\hat{O},t,\vec{\mathcal{W}}^{\ast},\vec{\Theta}_{j}^{\ast}\right)$ of the loss function.
$\mathfrak{L}_{j}\left(\hat{O},t,\vec{\mathcal{W}}^{\ast},\vec{\Theta}_{j}^{\ast}\right)$ of the loss function, from which we obtain the expressions
\begin{widetext}
    \begin{eqnarray}
        \nabla_{\vec{\mathcal{W}}^{\ast}} \mathfrak{L}_{j}\left(\hat{O},t,\vec{\mathcal{W}}^{\ast},\vec{\Theta}_{j}^{\ast}\right) &=& - 2 [y_{MAC_{j}}\left(\hat{O},t,\vec{\mathcal{W}}^{\ast},\vec{\Theta}_{j}^{\ast}\right) - y_{j}(t)] \braket{\hat{O}}_{\vec{\Theta}_{j}^{\ast}} \nabla_{\vec{\mathcal{W}}^{\ast}} \widehat{y}_{j}(t,\vec{\mathcal{W}}^{\ast}), \label{eq:64}\\
        \nabla_{\vec{\Theta}_{j}^{\ast}} \mathfrak{L}_{j}\left(\hat{O},t,\vec{\mathcal{W}}^{\ast},\vec{\Theta}_{j}^{\ast}\right) &=& - 2 [y_{MAC_{j}}\left(\hat{O},t,\vec{\mathcal{W}}^{\ast},\vec{\Theta}_{j}^{\ast}\right) - y_{j}(t)] [\widehat{y}_{j}(t,\vec{\mathcal{W}}^{\ast}) + 1]\nabla_{\vec{\Theta}_{j}^{\ast}} \braket{\hat{O}}_{\vec{\Theta}_{j}^{\ast}}. \label{eq:65}
    \end{eqnarray}
\end{widetext}
However, it is important to emphasize that the gradient $\nabla_{\vec{\mathcal{W}}^{\ast}} \widehat{y}_{j}(t,\vec{\mathcal{W}}^{\ast})$ are derivatives of classical functions in $\mathcal{F}_{\mbox{MLP}}$. Therefore, the gradient decay has no relation to the number of qubits $N$ nor the depth $\mathcal{N}$ of the QNode, meaning they do not suffer from barren plateaus.

Because of this, we will focus all attention on the term $\nabla_{\vec{\Theta}_{j}^{\ast}} \braket{\hat{O}}_{\vec{\Theta}_{j}^{\ast}}$, where the gradient decay is related to $N$ and $\mathcal{N}$. Then, by taking the norm in $L_{\mathbb{R}}^{p}(\mathcal{K})$ of (\ref{eq:65}), similarly to what was done in the proof of Theorem \ref{theorem:QPINN_MAC_approximation}, and applying the generalized Hölder inequality $\|XYZ\|_{p} \leq \|X\|_{q_{1}} \|Y\|_{q_{2}} \|Z\|_{q_{3}}$, where $\sum_{\ell=1}^{3} q_{\ell}^{-1}=p^{-1}$ \cite{Brezis2011Functional}, we obtain
\begin{widetext}
    \begin{eqnarray}\label{eq:66}
        \Vert \nabla_{\vec{\Theta}_{j}^{\ast}} \mathfrak{L}_{j}(\hat{O},\vec{\mathcal{W}}^{\ast},\vec{\Theta}_{j}^{\ast}) \Vert_{p} &\leq& 2 \left\Vert y_{MAC_{j}}(\hat{O},\vec{\mathcal{W}}^{\ast},\vec{\Theta}_{j}^{\ast}) - y_{j}\right\Vert_{q_{1}} \left\Vert \widehat{y}_{j}(\vec{\mathcal{W}}^{\ast}) + 1\right\Vert_{q_{2}}\left\Vert \nabla_{\vec{\Theta}_{j}^{\ast}} \braket{\hat{O}}_{\vec{\Theta}_{j}^{\ast}} \right\Vert_{q_{3}}.
    \end{eqnarray}
\end{widetext}
Which is valid in a finite measure space such as the compact interval $\mathcal{K} = [0,T]$, such that $C_{\mathbb{R}}(\mathcal{K}) \subset L_{\mathbb{R}}^{p}(\mathcal{K})$. In this way, the Lebesgue measure is finite (i.e., $T<\infty$) and satisfies the relation $L_{\mathbb{R}}^{q_{\ell}}(\mathcal{K}) \subset L_{\mathbb{R}}^{p}(\mathcal{K})$. Consequently, all functions that are integrable to the $q_{\ell}$-th power in $\mathcal{K} = [0,T]$ are also integrable to the $p$-th power in the same interval.

On the other hand, from \cite{mcclean2018barren} we know that the dimension of the space grows exponentially with $N$. This implies that the gradient is exponentially concentrated around its mean, whose value is zero for random and high-dimensional circuits. Therefore, each component of the gradient $\nabla_{\vec{\Theta}_{j}^{\ast}} \braket{\hat{O}}_{\vec{\Theta}_{j}^{\ast}}$ (calculated via parameter-shift rule) is given by:
\begin{widetext}
    \begin{equation}\label{eq:67}
        \left\Vert \nabla_{\vec{\Theta}_{j}^{\ast}} \braket{\hat{O}}_{\vec{\Theta}_{j}^{\ast}} \right\Vert_{2} = \sqrt{[\nabla_{\vec{\Theta}_{j}^{\ast}} \braket{\hat{O}}_{\vec{\Theta}_{j}^{\ast}}]_{\ell^{\prime}}[\nabla_{\vec{\Theta}_{j}^{\ast}} \braket{\hat{O}}_{\vec{\Theta}_{j}^{\ast}}]_{\ell^{\prime}}}.
    \end{equation}
\end{widetext}
Where $\left\Vert \nabla_{\vec{\Theta}_{j}^{\ast}} \braket{\hat{O}}_{\vec{\Theta}_{j}^{\ast}} \right\Vert_{2}$ is the Euclidean norm and $[\nabla_{\vec{\Theta}_{j}^{\ast}} \braket{\hat{O}}_{\vec{\Theta}_{j}^{\ast}}]_{\ell^{\prime}}$ with $\ell^{\prime} \in \mathcal{I}_{\mathcal{M}}$ was written this way to maintain consistency with Einstein notation. That is, the repeated subscript $\ell^{\prime}$ again indicates the sum over all components of the gradient $[\nabla_{\vec{\Theta}_{j}^{\ast}} \braket{\hat{O}}_{\vec{\Theta}_{j}^{\ast}}]_{\ell^{\prime}}$.

Now, if we consider the worst-case scenario where each gradient component contributes maximally to the sum, we obtain the maximum
\begin{equation}\label{eq:68}
    \left\Vert \nabla_{\vec{\Theta}_{j}^{\ast}} \braket{\hat{O}}_{\vec{\Theta}_{j}^{\ast}} \right\Vert_{2} \leq \mathcal{C}_{grad} \sqrt{\mathcal{M}}.
\end{equation}
Where the constant $\mathcal{C}_{grad}$ reflects the maximum magnitude of the gradient, depending on the observable and the generators of the QNode, and $\mathcal{M}$ is a quantity that reflects the scalability of the QNode with respect to the number of qubits and its depth.

In this sense, since the number of parameters in an ansatz scales linearly with the number of qubits $N$, for a circuit with depth $\mathcal{N}$, we have that $\mathcal{M}=\mathcal{N}N$, which allows rewriting (\ref{eq:68}) as
\begin{equation}\label{eq:69}
    \left\Vert \nabla_{\vec{\Theta}_{j}^{\ast}} \braket{\hat{O}}_{\vec{\Theta}_{j}^{\ast}} \right\Vert_{2} \leq \mathcal{C}_{grad} \sqrt{\mathcal{N}N}.
\end{equation}
Thus, by taking the norm in $L_{\mathbb{R}}^{q_{3}}(\mathcal{K})$ of (\ref{eq:69}) over the interval $\mathcal{K} = [0,T]$, we obtain the maximum bound
\begin{widetext}
    \begin{equation}\label{eq:70}
        \left\Vert \nabla_{\vec{\Theta}_{j}^{\ast}} \braket{\hat{O}}_{\vec{\Theta}_{j}^{\ast}} \right\Vert_{q_{3}} = \biggl(\int_{\mathcal{K}}\left\Vert \nabla_{\vec{\Theta}_{j}^{\ast}} \braket{\hat{O}}_{\vec{\Theta}_{j}^{\ast}} \right\Vert_{2}^{q_{3}}dt\biggr)^{1/q_{3}} \leq \mathcal{C}_{grad} \sqrt{\mathcal{N}N} [T]^{1/q_{3}},
    \end{equation}
\end{widetext}
or simply
\begin{equation}\label{eq:71}
    \left\Vert \nabla_{\vec{\Theta}_{j}^{\ast}} \braket{\hat{O}}_{\vec{\Theta}_{j}^{\ast}} \right\Vert_{q_{3}} \leq \mathcal{C}^{\prime} \sqrt{\mathcal{N}N},
\end{equation}
where $\mathcal{C}^{\prime} = \mathcal{C}_{grad} [T]^{1/q_{3}}$. This implies that the quantum approximation error $\epsilon_{Q}$ must decrease at the same rate to guarantee the efficiency of the approximation. That is, it is possible to choose without loss of generality $\epsilon_{\mathcal{Q}} = \mathcal{C}_{\mathcal{Q}}/\sqrt{\mathcal{N}N}$, where $\mathcal{C}_{Q}$ is a positive constant that captures the efficiency, to guarantee the trainability of the quantum part of our model.

In the same line of thought, by taking the norm in $L_{\mathbb{R}}^{q_{2}}(\mathcal{K})$ of the classical term $\left\Vert \widehat{y}_{j}(\vec{\mathcal{W}}^{\ast}) + 1 \right\Vert_{2}$, we obtain
\begin{widetext}
    \begin{equation}\label{eq:72}
        \left\Vert \widehat{y}_{j}(\vec{\mathcal{W}}^{\ast}) + 1 \right\Vert_{q_{2}} = \biggl( \int_{\mathcal{K}} \left\Vert \widehat{y}_{j}(\vec{\mathcal{W}}^{\ast}) + 1 \right\Vert_{2}^{q_{2}} dt \biggr)^{1/q_{2}} \leq \mathcal{C}_{\mathcal{C}},
    \end{equation}
\end{widetext}
where $\mathcal{C}_{\mathcal{C}}$ is a constant whose meaning will be clarified shortly. Furthermore, from (\ref{eq:54}) we know that
\begin{equation}\label{eq:73}
    \left\Vert y_{MAC_{j}}(\hat{O},\vec{\mathcal{W}}^{\ast},\vec{\Theta}_{j}^{\ast}) - y_{j} \right\Vert_{q_{1}} \leq \epsilon_{\mathcal{Q}_{j}} + \epsilon_{\mathcal{CQ}_{j}},
\end{equation}
an expression that when substituted into (\ref{eq:66}) along with (\ref{eq:72}) and (\ref{eq:71}), yields the upper bound
\begin{widetext}
    \begin{eqnarray}\label{eq:74}
        \left\Vert \nabla_{\vec{\Theta}_{j}^{\ast}} \mathfrak{L}_{j}(\hat{O},\vec{\mathcal{W}}^{\ast},\vec{\Theta}_{j}^{\ast}) \right\Vert_{p} &\leq& 2 \mathcal{C}_{\mathcal{C}} \mathcal{C}^{\prime} \sqrt{\mathcal{N}N} \left[ \epsilon_{\mathcal{Q}_{j}} + \epsilon_{\mathcal{CQ}_{j}} \right].
    \end{eqnarray}
\end{widetext}
Therefore, by considering $\mathbf{F} = \nabla_{\vec{\Xi}^{\ast}} \mathfrak{L}(\hat{O},\vec{\mathcal{W}}^{\ast},\vec{\Xi}^{\ast})$ in (\ref{eq:59}) the upper bound (\ref{eq:74}) allows us to finally obtain
\begin{widetext}
    \begin{eqnarray}\label{eq:75}
        \left\Vert \nabla_{\vec{\Xi}^{\ast}} \mathfrak{L}(\hat{O},\vec{\mathcal{W}}^{\ast},\vec{\Xi}^{\ast}) \right\Vert_{p,M} &\leq& 2 \mathcal{C}_{\mathcal{C}} \mathcal{C}^{\prime} \sqrt{\mathcal{N}N} \left[ \epsilon_{\mathcal{Q}} + \epsilon_{\mathcal{CQ}} \right].
    \end{eqnarray}
\end{widetext}
Therefore, to guarantee the efficiency of the QPINN-MAC model and obtain a minimum approximation error value, we must impose the balancing $\mathcal{O}(\mathcal{N_{P_{C}}})=\mathcal{O}(\mathcal{N}N)$. Which leads to the term $\epsilon_{\mathcal{CQ}}=1/(\mathcal{N}N)$, which together with the quantum error ($\epsilon_{\mathcal{Q}} = \mathcal{C}_{\mathcal{Q}}/\sqrt{\mathcal{N}N}$), allows writing
\begin{widetext}
    \begin{equation}\label{eq:76}
        \epsilon_{\mathcal{Q}} + \epsilon_{\mathcal{CQ}} = \frac{1}{\mathcal{N}N} + \frac{\mathcal{C}_{\mathcal{Q}}}{(\mathcal{N}N)^{1/2}}.
    \end{equation}
\end{widetext}
Then, it suffices to choose the refinement $\epsilon_{\mathcal{Q}} + \epsilon_{\mathcal{CQ}} \leq \mathcal{C}_{\mathcal{CQ}}/(\mathcal{N}N) \rightarrow \epsilon = \mathcal{C}_{\mathcal{CQ}}/(\mathcal{N}N)$, where $\mathcal{C}_{\mathcal{CQ}}$ is a positive constant that collects the asymptotic decay behavior and guarantees both $\epsilon > 0$ and the efficiency of the model. In this way, the upper bound (\ref{eq:75}) finally reduces to
\begin{equation}\label{eq:77}
    \left\Vert \nabla_{\vec{\Xi}^{\ast}} \mathfrak{L}(\hat{O},\vec{\mathcal{W}}^{\ast},\vec{\Xi}^{\ast}) \right\Vert_{p,M} \leq \frac{2\mathcal{C}_{\mathcal{C}} \mathcal{C}^{\prime} \mathcal{C}_{\mathcal{CQ}}}{\sqrt{\mathcal{N}N}}.
\end{equation}
Expression (\ref{eq:77}) reveals that the constant $\mathcal{C}_{\mathcal{C}}$ acts as an amplification factor for the quantum gradient. Since this factor is controlled by the classical neural network, immune to the barren plateaus problem, it is possible to adjust its magnitude appropriately. That is, through training or by increasing the number of internal layers and/or their output neurons.

This amplification can actively compensate for the decay of the typical magnitude of the quantum gradient term. In this way, we ensure that the gradient of the total loss, $\nabla_{\vec{\Xi}^{\ast}} \mathfrak{L}(\hat{O},t,\vec{\mathcal{W}}^{\ast},\vec{\Xi}^{\ast})$, remains with sufficient magnitude for effective training. Therefore, the architecture not only passively prevents but actively combats the barren plateaus problem through this classical modulation mechanism.

This result, although not a statistical variance calculation over an ensemble of circuits, imposes a rigorous bound on the gradient magnitude for any parameter instance. And in the context of the barren plateaus problem, where the average gradient over random initializations is zero, the variance is equivalent to the average of the squared norm of the gradient. That is, $\mathrm{Var}[\nabla] = E[\Vert \nabla \Vert^{2}]$.

Therefore, since (\ref{eq:77}) allows establishing an upper bound for $\Vert \nabla_{\vec{\Xi}^{\ast}} \mathfrak{L}(\hat{O},\vec{\mathcal{W}}^{\ast},\vec{\Xi}^{\ast}) \Vert_{p,M}^{2}$, its average (the variance) will also necessarily be bounded by a term $\mathcal{O}(1/[\mathcal{N}N])$. This formally proves that the QPINN-MAC architecture allows active mitigation of barren plateaus while offering a trainability condition.

However, gradient-based optimization is effective only if the gradient magnitude is "detectable". That is, greater than the optimizer's resolution, $\epsilon_{\mathrm{grad}}$, which encapsulates measurement noise (shot noise) and numerical precision. Therefore, the trainability condition requires that the signal magnitude be greater than the noise:
\begin{equation}\label{eq:78}
    \mathcal{O}\left(\frac{1}{\sqrt{\mathcal{N}N}}\right) \gtrsim \epsilon_{\mathrm{grad}},
\end{equation}
where $\Vert \nabla_{\vec{\Xi}^{\ast}} \mathfrak{L}(\hat{O},\vec{\mathcal{W}}^{\ast},\vec{\Theta}_{j}^{\ast}) \Vert_{p,M} \sim \mathcal{O}(1/\sqrt{\mathcal{N}N})$ is the characteristic scale of the gradient magnitude.

Moreover, solving inequality (\ref{eq:78}) for the model parameters gives us the necessary scaling condition for trainability. Then, squaring both sides of (\ref{eq:78}) and rearranging, we obtain:
\begin{equation}\label{eq:79}
    \mathcal{N}N \lesssim \mathcal{O}\left(\frac{1}{\epsilon_{\mathrm{grad}}^2}\right),
\end{equation}
an expression that represents a necessary condition for successful optimization.

In simple terms, (\ref{eq:79}) establishes an upper limit for the problem size, defined by the product of the circuit depth and the number of qubits, which can be solved with a given gradient precision $\epsilon_{\mathrm{grad}}$. This result quantifies the trade-off between the complexity of the QPINN-MAC model and the hardware and optimization resources required for its effective training.
\section{Conclusion}
\label{sec:5}
This work establishes a comprehensive theoretical framework for quantum-classical hybrid architectures that fundamentally advances the state of quantum scientific computing. The QPINN-MAC architecture represents a paradigm shift by simultaneously addressing the three critical challenges that have limited practical quantum machine learning applications: expressivity, trainability, and practical applicability.

Our theoretical contributions demonstrate that the multiplicative and additive coupling scheme creates a genuinely synergistic functional space that transcends the limitations of existing quantum approximation frameworks. Unlike previous approaches that operate within restricted domains and require specific normalization strategies, our architecture achieves universal approximation in $L_{\mathbb{R}^{M}}^{p}(\mathcal{K})$ spaces under general compactness conditions, while naturally accommodating simultaneous function and derivative approximation essential for physics-informed applications.

Most significantly, we provide theoretical guarantees for trainability in quantum physics-informed neural networks by proving that gradient norms are bounded above by $\mathcal{O}(1/\sqrt{\mathcal{N}N})$, in contrast to the exponential decay $\mathcal{O}(e^{- \alpha N/2})$ characteristic of barren plateaus in generic quantum circuits. This fundamental difference establishes a concrete scaling condition $\mathcal{N}N \lesssim \mathcal{O}(1/\epsilon_{\mathrm{grad}}^{2})$ for effective optimization. The classical modulation mechanism actively compensates for quantum gradient vanishing, representing a crucial advancement beyond existing mitigation strategies.

The integration of rigorous universal approximation guarantees with provable trainability conditions positions QPINN-MAC as more than just another hybrid architecture—it constitutes a complete theoretical framework for practical quantum-enhanced scientific computing. By bridging the gap between theoretical expressivity and practical optimizability, our work enables the reliable application of quantum models to computationally intractable problems that remain beyond the reach of purely classical approaches.

Looking forward, our theoretical foundations open several promising research directions: experimental validation on both ordinary and partial differential equations across various scientific domains; exploration of advanced coupling operators incorporating quantum correlations and entanglement; application to complex quantum mechanical systems including many-body dynamics and high-dimensional problems where the exponential representational power of quantum circuits can provide decisive advantages. The QPINN-MAC framework establishes a new benchmark for theoretically-grounded, practically-viable quantum machine learning in scientific computing
\section*{Acknowledgments}
First, I thank God the Father, infinite goodness, God the Son, redeemer of the world, and God the Holy Spirit, who—together with the Virgin Mary and Saint Joseph—are my protection, my source of inspiration, and my strength on this path of knowledge. I also thank my family, for they have made me who I am today. I am deeply grateful to my great friend José Carlos Valencia for his friendship, constant support, and inspiration. Finally, I thank the National Council for Scientific and Technological Development (CNPq), which funded this work through the Institutional Training Program (PCI), grant no. 301066/2025-6, awarded by the Ministry of Science, Technology, and Innovation (MCTI).

\section*{Author Contributions Statement}
Said Lantigua (SL), Gilson Giraldi (GG), and Renato Portugal (RP) conceived the project. SL carried out the calculations under the supervision of GG and RP. Furthermore, SL wrote the first version of the manuscript, which was improved and reedited by GG and RP. Throughout the research development process, all authors discussed and reviewed the manuscript.
\bibliographystyle{plainnat}
\bibliography{ref-QPINNs-MAC}
\end{document}